\documentclass[11pt]{article}
\usepackage{amsmath}
\usepackage{amssymb}
\usepackage{amsfonts}
\usepackage{amsthm}
\usepackage{bm}
\usepackage{mathtools}
\mathtoolsset{showonlyrefs} 
\usepackage{geometry}
\usepackage{setspace}
\usepackage{graphicx}
\usepackage{booktabs}
\usepackage{tabularx}
\usepackage{subcaption}
\usepackage[round,authoryear]{natbib}
\usepackage{xcolor}
\usepackage{xr-hyper} 
\usepackage{hyperref}
\makeatletter
\providecommand{\frozenref}[2]{\expandafter\gdef\csname frozen@#1\endcsname{#2}}
\providecommand{\xref}[1]{\@ifundefined{r@#1}{\@ifundefined{frozen@#1}{\textbf{??}}{\csname frozen@#1\endcsname}}{\ref{#1}}}

\makeatother
\frozenref{S-alg:MHG}{S1}
\frozenref{S-fig:sim-traces}{S1}
\frozenref{S-sec:supp_app}{S4}
\frozenref{S-sec:supp_mcmc}{S2}
\frozenref{S-sec:supp_mcmc_gev}{S2.3}
\frozenref{S-sec:supp_mcmc_gpd}{S2.4}
\frozenref{S-sec:supp_moves}{S2.6}
\frozenref{S-sec:supp_rv}{S1}
\frozenref{S-sec:supp_sim}{S3}
\frozenref{S-tab:sim-rep1}{S4}
\usepackage{authblk}

\hypersetup{
  colorlinks=true,
  linkcolor=blue, 
  citecolor=blue, 
  urlcolor=blue,  
  filecolor=blue
}
\newcommand{\keywords}[1]{%
  \par\noindent\textbf{Keywords:} #1\par
}

\usepackage[ruled,vlined]{algorithm2e}
\usepackage{float}
\newtheorem{theorem}{Theorem}[subsection]
\newtheorem{proposition}[theorem]{Proposition}
\newtheorem{lemma}[theorem]{Lemma}

\theoremstyle{remark}
\newtheorem{remark}[theorem]{Remark}

\newenvironment{proof*}{\begin{proof}}{\end{proof}}
\usepackage{adjustbox}
\title{Tail persistence of extremes at finite and asymptotic levels with Bayesian hierarchical inference}

\author[1]{Nikita Rawat}
\author[2]{Shiv Kumar Yadav}
\author[1]{Dreamlee Sharma}
\author[3]{Rishikesh Yadav}
\affil[1]{Department of Statistics, University of Delhi}
\affil[2]{ Indian Statistical Institute, Delhi Centre}
\affil[3]{School of Mathematical and Statistical Sciences, Indian Institute of Technology Mandi}

\date{\today}

\begin{document}
\maketitle
\noeqref{eq:boundedXiLink,eq:fullHierarchy,eq:gevCDF,eq:gevPYdraw,eq:gpdCDF,eq:gpdPYdraw,eq:gpdPZdraw,eq:persistence,eq:posteriorPersistenceDraw,eq:tildeLinkGeneral,eq:tildeLinkXiGeneral,eq:xiRecover}
\begin{abstract}
Tail persistence characterizes the relative escalation of observations beyond an extreme threshold and provides a complementary perspective to conventional summaries such as return levels and tail indices. Existing literature is mainly distribution-based and does not account for uncertainty, nonstationarity, or covariate effects. We develop a Bayesian framework for modeling tail persistence under generalized extreme value and generalized Pareto models under regular variation assumptions. We first make explicit the connection between tail persistence and the extreme value index under regular variation, which identifies the asymptotic persistence with $1/(1-\xi)$ and links persistence to tail heaviness. We then derive closed-form expressions for persistence at finite quantile levels under both models and show that, for any location--scale extreme value family with a common shape parameter, finite-level persistence depends on the location and scale parameters only through their ratio. Motivated by these results and by the weak identifiability of observation-specific shape parameters, we formulate a Bayesian hierarchical extreme value model in which the shape parameter is held constant over time, while the location and scale parameters vary with covariates and latent effects. Under this formulation the asymptotic persistence is a single, well-identified quantity, whereas finite-level persistence evolves over time through the location--scale structure. The framework provides posterior inference for both quantities together with credible intervals and probabilistic assessment of temporal evolution and covariate effects. Through simulation studies and real-data applications, we demonstrate that persistence provides an interpretable characterization of the escalation of extremes beyond high thresholds, complementing classical extreme value summaries while enabling comprehensive uncertainty quantification in nonstationary settings. The methodology is implemented in an open-source \texttt{R} package. 
\end{abstract}

\keywords{Bayesian inference; extreme value theory; tail persistence; generalized extreme value; generalized Pareto; nonstationarity}

\section{Introduction}
\label{sec:intro}

Assessing the risk posed by environmental, hydrological and financial extremes requires knowing not only how often a high level is exceeded, but also by how much it is exceeded once it has been reached. Extreme value theory provides asymptotically justified models for both questions: the generalized extreme value (GEV) distribution for block maxima and the generalized Pareto distribution (GPD) for threshold exceedances \citep{Fisher1928,Gnedenko1943,Balkema1974,Pickands1975,Davison1990,Coles2001,deHaan2006}. Inference based on these models is usually summarized through the shape parameter $\xi$, return levels and exceedance probabilities. These quantities describe the frequency and the magnitude of extremes. They do not describe, in a directly interpretable and scale-free way, the \emph{relative escalation} of observations beyond a level that is already extreme. Two rainfall series may share the same $100$-year return level, yet exceedances of that level may stay close to it in one series and overshoot it substantially in the other. The two situations have different consequences for, e.g., the design of flood defences.

A natural summary of this escalation is the \emph{tail persistence}
\[
P(u)\;=\;\frac{E\{Y\mid Y>Q(u)\}}{Q(u)}\;=\;\frac{1}{(1-u)\,Q(u)}\int_u^1 Q(s)\,ds,
\qquad Q(u)>0,
\]
where $Q$ is the quantile function of $Y$. Thus $P(u)-1$ is the mean excess beyond the $u$th quantile, measured in units of that quantile. In this quantile form, $P$ was recently proposed by \citet{Sankaran2026a} as a descriptive, distribution-level measure of upper-tail growth. The same ratio is familiar in quantitative risk management as the ratio of expected shortfall (tail conditional expectation) to value-at-risk at level $u$ \citep{AcerbiTasche2002,McNeil2015}, and it is closely related to the mean residual life function used for threshold selection \citep{Davison1990}. If the tail of $Y$ is regularly varying with tail index $0<\xi<1$, Karamata's theorem gives $P(u)\to1/(1-\xi)$ as $u\uparrow1$. This limit underlies second-order expansions of tail conditional expectations and the estimation of extreme risk measures \citep{HuaJoe2011,ElMethni2014,Daouia2018}.

Three aspects of persistence have nevertheless received little attention. First, the limit $1/(1-\xi)$ describes the infinitely remote tail. It equals one throughout the Gumbel and Weibull domains of attraction, and it says nothing about escalation at the finite levels, such as $u=0.99$, at which risk is actually assessed. Closed-form expected shortfalls are available for the GPD \citep{McNeil2015}. However, how finite-level persistence depends on the location and scale parameters of a nonstationary extreme value model has not been characterized or exploited. Second, existing treatments of persistence are descriptive: they quantify neither its estimation uncertainty nor its dependence on time or covariates. Third, it is unclear how persistence should be modelled when the data are nonstationary. Letting the shape parameter vary with time would make the asymptotic persistence time-varying. But $\xi$ is the most poorly identified parameter of extreme value models \citep{Coles2001,Papalexiou2013}, and observation-specific shape parameters, each informed by a single maximum or exceedance, are determined essentially by the prior.

Regression and hierarchical models for nonstationary extremes are well established. Frequentist approaches link the GEV or GPD parameters to covariates, usually letting the location and scale vary while the shape is held fixed \citep{Davison1990,Coles2001,ChavezDemoulin2005,EastoeTawn2009}. Bayesian formulations \citep{ColesTawn1996,Castellanos2007,Nascimento2012,Nascimento2017,Villa2017} propagate parameter uncertainty to derived quantities and accommodate latent temporal or spatial effects \citep{Cooley2007,Opitz2018,Yadav2021,CastroCamilo2022}. In this literature, the inferential targets are the distributional parameters and return levels. Persistence, when it is considered at all, is a by-product of a fitted model, not a quantity whose structure guides how the model is specified.

In this paper, we make tail persistence the primary target of inference in nonstationary extreme value models. The starting point is a structural property. For any location--scale family $Y=\mu+\sigma W$ in which the standardized variable $W$ depends only on a shape parameter $\xi$, persistence at every finite level depends on $(\mu,\sigma)$ only through the ratio $\lambda=\mu/\sigma$, and it is strictly decreasing in $\lambda$. A model with a common shape and time-varying location and scale therefore separates two quantities:
\begin{itemize}
\item a single, time-invariant \emph{asymptotic persistence} $P_\infty=1/(1-\xi)$, informed by all observations;
\item a time-varying \emph{finite-level persistence} $P_i(p)$, driven by $\lambda_i=\mu_i/\sigma_i$ and hence identified through the location and scale channels, which are informed by the bulk of the extremal data.
\end{itemize}
Holding $\xi$ constant thus removes the identifiability problem without removing nonstationarity from persistence. Our contributions are as follows.
\begin{enumerate}
\item We derive exact closed-form expressions for finite-level persistence under the GPD and GEV models (Propositions~\ref{prop:gpdPersistence} and~\ref{prop:gevPersistence}). They rest on a mean-excess representation (Lemma~\ref{lem:meanExcess}) and on a location--scale reduction (Proposition~\ref{prop:locScale}), and they connect to the asymptotic persistence of Theorem~\ref{thm:persistence}, including its degeneration to one for $\xi\le0$. Under the GPD, persistence of the excesses does not depend on the scale at any level, whereas persistence of the levels depends on the threshold-to-scale ratio. Under the GEV, the location--scale ratio has a direct interpretation through the lower endpoint of the distribution.
\item We show that, under a common shape, the posterior probability that persistence at time $i$ exceeds that at time $i'$ equals $\Pr(\lambda_i<\lambda_{i'}\mid\mathbf y)$. The same probability therefore holds at every level $p$; see~\eqref{eq:persistenceContrast}.
\item We develop Bayesian hierarchical GEV and GPD models with a common shape and with location and scale driven by covariates and latent effects. The shape is linked through a bounded map, which guarantees finite persistence a posteriori. In the sampler, all regression coefficients and variance components are updated by conjugate Gibbs draws and the common shape by elliptical slice sampling \citep{murray2010elliptical}. Probability-integral-transform-preserving moves address the posterior dependence between $\xi$ and the scale parameters.
\item We study the frequentist properties of the procedure by simulation and apply it to monthly England and Wales precipitation, 1950--2025. Both models place the upper tail in the Weibull domain, so $P_\infty=1$. Finite-level persistence nevertheless remains informative, and it evolves in opposite directions for annual maxima and for threshold exceedances. The methodology is implemented in an open-source \texttt{R} package.
\end{enumerate}

The paper is organized as follows. Section~\ref{sec:background} recalls the extreme value models and defines tail persistence. Section~\ref{sec:methodology} presents the asymptotic and finite-level theory of persistence, the Bayesian hierarchical framework, and its GPD and GEV specializations. Section~\ref{sec:simulation} reports the simulation study, Section~\ref{sec:application} the application, and Section~\ref{sec:conclusion} concludes. Proofs of all theoretical results are given in Appendix~\ref{app:proofs}. The Supplementary Material contains background on regular variation (Section~\xref{S-sec:supp_rv}), the details of posterior computation, namely the full conditional distributions, the complete sampler and the additional MCMC moves (Section~\xref{S-sec:supp_mcmc}), and additional simulation and application results (Sections~\xref{S-sec:supp_sim}--\xref{S-sec:supp_app}); its sections, equations, tables, figures and algorithms are prefixed by ``S''.

\section{Background}
\label{sec:background}

\subsection{Extreme value models and regular variation}
\label{subsec:evModels}
\label{subsec:pot_details}
Let $Y_1,\ldots,Y_n$ be independent with common distribution function $F$ and upper endpoint $x_F=\sup\{x:F(x)<1\}$, and let $M_n=\max(Y_1,\ldots,Y_n)$. If there exist constants $a_n>0$ and $b_n\in\mathbb{R}$ such that $\Pr\{(M_n-b_n)/a_n\le x\}\to G(x)$ for a non-degenerate $G$, then $G$ belongs to the generalized extreme value (GEV) family \citep{Fisher1928,Gnedenko1943,deHaan2006},
\begin{align}
G_{\mu,\sigma,\xi}(x)
=
\exp\left\{
-\left[1+\xi\left(\frac{x-\mu}{\sigma}\right)\right]^{-1/\xi}
\right\},
\qquad 1+\xi(x-\mu)/\sigma>0,
\label{eq:gevCDF}
\end{align}
with location $\mu\in\mathbb{R}$, scale $\sigma>0$ and shape $\xi\in\mathbb{R}$; the cases $\xi>0$, $\xi=0$ (interpreted as the limit $\xi\to0$) and $\xi<0$ correspond to the Fr\'echet, Gumbel and Weibull domains of attraction. Under the same condition, the excess distribution $F_v(z)=\Pr(Y-v\le z\mid Y>v)$, $z\ge0$, is approximated as $v\to x_F$ by the generalized Pareto distribution (GPD) \citep{Balkema1974,Pickands1975},
\begin{equation}
G_{\sigma,\xi}(z)=
\begin{cases}
1-\left(1+\xi z/\sigma\right)^{-1/\xi}, & \xi\neq0,\\[1mm]
1-\exp(-z/\sigma), & \xi=0,
\end{cases}
\label{eq:gpdCDF}
\end{equation}
with the same shape parameter $\xi$. The block maxima approach fits~\eqref{eq:gevCDF} to maxima over blocks such as years, whereas the peaks-over-threshold (POT) approach fits~\eqref{eq:gpdCDF} to the excesses over a high threshold and therefore uses more of the extreme data; neither is uniformly preferable \citep{bucher2021horse}.

The Fr\'echet domain is characterized by regular variation: $F$ is in the maximum domain of attraction of the GEV with $\xi>0$ if and only if $\bar F=1-F$ is regularly varying with index $-1/\xi$, written $\bar F\in RV_{-1/\xi}$, that is,
\(
\lim_{x\to\infty}\bar F(tx)/\bar F(x)=t^{-1/\xi}
\)
for all $t>0$, or equivalently $\bar F(x)=x^{-1/\xi}L(x)$ for a slowly varying function $L$ \citep{bingham1989regular,deHaan2006,Resnick2007}. The tail index $\xi$ then measures tail heaviness, with larger $\xi$ corresponding to heavier tails. Formal definitions, standard examples and the inversion result for the quantile function used in Section~\ref{sec:methodology} are collected in Section~\xref{S-sec:supp_rv} of the Supplementary Material.

\subsection{Tail persistence}
\label{subsec:def_tail_pst}
Following \citet{Sankaran2026a}, who introduced it as the \emph{quantile-based effectiveness persistence function}, we measure upper-tail growth through a quantile-based ratio. Let
\(
Q(u)=F^{-1}(u), 0<u<1,
\)
denote the quantile function of $F$. The persistence function is defined as
\begin{equation}
P(u)
=
\frac{1}{(1-u)Q(u)}
\int_u^1Q(p)\,dp,
\qquad 0<u<1,
\label{eq:persistence}
\end{equation}
for those $u$ at which $Q(u)>0$ and the integral is finite.
The numerator represents the average quantile beyond the threshold $Q(u)$, while normalization by $Q(u)$ yields the average relative magnitude of observations exceeding the $u$th quantile.

Persistence complements conventional extreme-value summaries by characterizing the behavior of the upper tail beyond a prescribed threshold. For example, two regions may possess the same 100-year rainfall return level, yet substantially different persistence if exceedances beyond that level remain close to the threshold in one region but are considerably larger in the other. Several properties follow immediately from \eqref{eq:persistence}: 
\begin{itemize}
\item[(i)] since $Q(p)\ge Q(u)>0$ for $p\ge u$, it follows that $P(u)\ge1$, with equality only for degenerate upper tails; 
\item[(ii)] larger values of $P(u)$ indicate stronger upper-tail growth beyond the threshold, whereas values close to one imply relatively mild tail escalation; 
\item[(iii)] $\{P(u):u_0<u<1\}$ forms a persistence profile, providing a continuous characterization of tail behavior across a range of high quantiles rather than at a single return level or exceedance probability.
\end{itemize}

\section{Bayesian hierarchical persistence at asymptotic and finite quantile level}
\label{sec:methodology}
In this section, we develop a Bayesian framework for inference on tail persistence of extreme value models. We begin in Section \ref{subsec:Asymtot_tailPersistence} by establishing a theoretical connection between the persistence function and the tail index under regular variation, showing that the asymptotic persistence is uniquely determined by the extreme value index. In Section \ref{subsec:finitePersistence}, we derive exact expressions for persistence at finite quantile levels, first for a general location--scale family with a fixed shape parameter, and then explicitly for the generalized Pareto and generalized extreme value distributions. Motivated by these results, in Section \ref{subsec:generalBayesian_framework}, we introduce a general Bayesian hierarchical regression framework in which the shape parameter is common across observations, while the remaining distributional parameters vary with explanatory variables and latent effects; posterior inference is then developed directly for both the asymptotic and the finite-level persistence. Finally, in Section \ref{subsec:illust_example}, we specialize the methodology to the generalized Pareto and generalized extreme value models, and provide the corresponding interpretation of persistence in each case.

\subsection{Asymptotic characterization of tail persistence}
\label{subsec:Asymtot_tailPersistence}
The persistence function introduced in Section~\ref{subsec:def_tail_pst} provides a distribution-free characterization of upper-tail behavior through the quantile function. To develop a model-based inferential framework, it is necessary to establish its connection with classical extreme value theory. Theorem \ref{thm:persistence} shows that, under regular variation, the asymptotic persistence is completely determined by the extreme value index. Therefore, we may treat persistence as an alternative characterization of tail heaviness while retaining the theoretical foundation of extreme value models.

\begin{theorem}
\label{thm:persistence}
Let $Q(u)$ denote the quantile function of a distribution $F$, and let $P(u)$ be the persistence function as defined in \eqref{eq:persistence}.
Suppose the survival function $\bar F(x)=1-F(x)$ is regularly varying with tail index $\xi$, namely
\(
\bar F\in RV_{-1/\xi},
\, 0<\xi<1.
\) Then
\[
\lim_{u\uparrow1}P(u)
=
\frac{1}{1-\xi}.
\]
\end{theorem}
The proof, which applies Karamata's theorem to the regularly varying quantile function, is given in Appendix~\ref{app:proofThm}.
Theorem~\ref{thm:persistence} establishes a direct correspondence between persistence and the extreme value index. An immediate implication is that asymptotic persistence depends exclusively on the tail parameter $\xi$, independent of the location and scale of the underlying distribution. We denote this limit by
\begin{equation}
P_\infty(\xi) := \lim_{u\uparrow1}P(u) = \frac{1}{1-\xi},
\qquad 0<\xi<1,
\label{eq:Pinf}
\end{equation}
and refer to it as the \emph{asymptotic persistence}. Consequently, statistical inference for $P_\infty$ may be carried out through inference on $\xi$. As we show in Section~\ref{subsec:finitePersistence}, however, persistence at any \emph{finite} quantile level does depend on the location and scale parameters, and it is through these parameters that temporal and covariate variation in persistence is transmitted in the framework of Section~\ref{subsec:generalBayesian_framework}.

\begin{remark}[Persistence and regular variation]
\label{rem:regularVariationCaveat}
$P_\infty=1/(1-\xi)$ measures regular-variation tail heaviness only
when $\xi>0$ (Fr\'echet domain), since $\xi>0$ is equivalent to
$\bar F(x)\sim x^{-1/\xi}L(x)$ for slowly varying $L$. For $\xi=0$
(Gumbel domain), $P(u)\to1$ regardless of whether $F$ is heavy-tailed
(e.g.\ log-normal) or light-tailed (e.g.\ normal); for $\xi<0$ (Weibull
domain, permitted whenever $\xi_{\min}<0$), $F$ has a finite endpoint
$x_F$ and $P(u)\to1$, because $Q(p)\to x_F$ as $p\uparrow1$. The asymptotic
persistence is thus a valid regular-variation persistence measure
only where $\xi$ is credibly positive; therefore reporting
$\Pr(\xi>0\mid\mathbf y)$ alongside $P_\infty$ is recommended whenever this
interpretation is invoked. For $\xi\le0$, the general definition
$P_\infty := \lim_{u\uparrow1}P(u) = 1$ applies (Propositions~\ref{prop:gpdPersistence}--\ref{prop:gevPersistence} below), and the finite-level persistence of Section~\ref{subsec:finitePersistence} remains well defined and informative.
\end{remark}

\subsection{Persistence at finite quantile levels}
\label{subsec:finitePersistence}

In practice, persistence is assessed at high but finite levels, $p\in[p_0,1)$, such as $p=0.99$ or $p=0.999$. In this section we derive exact expressions for $P(p)$ under extreme value models. We first give a representation of $P(p)$ in terms of the quantile mean excess, then a location--scale reduction that applies to any family with a common shape parameter, and finally closed-form expressions for the GPD and GEV models.

\begin{lemma}[Mean-excess representation]
\label{lem:meanExcess}
Let $Y\sim F$, with $F$ continuous and strictly increasing on its support, $E|Y|<\infty$, and let $p\in(0,1)$ be such that $Q(p)>0$. Define the \emph{quantile mean} and the \emph{quantile mean excess}
\begin{equation}
m(p) := \frac{1}{1-p}\int_p^1 Q(s)\,ds,
\qquad
e(p) := m(p)-Q(p).
\label{eq:quantileMeanExcess}
\end{equation}
Then $m(p)=E[Y\mid Y>Q(p)]$, $e(p)=E[Y-Q(p)\mid Y>Q(p)]\ge0$, and
\begin{equation}
P(p) = \frac{m(p)}{Q(p)} = 1 + \frac{e(p)}{Q(p)}.
\label{eq:PmeanExcess}
\end{equation}
\end{lemma}
The proof of Lemma~\ref{lem:meanExcess} is given in Appendix~\ref{app:proofLemma}. Lemma~\ref{lem:meanExcess} shows that persistence is one plus the mean excess beyond the $p$th quantile, \emph{measured in units of that quantile}. In particular, $P(p)$ is a relative (dimensionless) quantity: it is invariant to multiplicative rescaling of $Y$, but not to shifts of $Y$.

\begin{remark}[Relation to expected shortfall]
\label{rem:ESVaR}
By Lemma~\ref{lem:meanExcess}, $P(p)=m(p)/Q(p)$ is the ratio of the expected shortfall (tail conditional expectation) $m(p)=E[Y\mid Y>Q(p)]$ to the value-at-risk $Q(p)$ at level $p$ \citep{AcerbiTasche2002,McNeil2015}, and Theorem~\ref{thm:persistence} is the classical limit of this ratio for regularly varying tails \citep{HuaJoe2011,ElMethni2014}. Our use of the ratio differs in two respects: it serves as a scale-free summary of tail escalation rather than as a risk measure, and its behaviour at \emph{finite} levels under location--scale extreme value families, established next, is exploited directly in the construction of nonstationary models.
\end{remark}

\begin{proposition}[Location--scale reduction]
\label{prop:locScale}
Let $W$ be a random variable with continuous, strictly increasing distribution function $F_W$ depending on a shape parameter $\xi$ only, with quantile function $Q_W(\cdot)$, quantile mean $m_W(\cdot)$ and quantile mean excess $e_W(\cdot)$ as in \eqref{eq:quantileMeanExcess}. Let $Y=\mu+\sigma W$ with $\mu\in\mathbb{R}$ and $\sigma>0$, and define the \emph{location--scale ratio}
\begin{equation}
\lambda := \frac{\mu}{\sigma}.
\label{eq:lambdaDef}
\end{equation}
Then, for every $p$ with $Q_Y(p)>0$, equivalently $\lambda+Q_W(p)>0$,
\begin{equation}
P_Y(p)
=
1 + \frac{e_W(p)}{\lambda + Q_W(p)}.
\label{eq:PlocScale}
\end{equation}
Consequently:
\begin{enumerate}
\item[(i)] $P_Y(p)$ depends on $(\mu,\sigma)$ only through $\lambda=\mu/\sigma$;
\item[(ii)] $P_Y(p)$ is strictly decreasing in $\lambda$ whenever $e_W(p)>0$, with
\begin{equation}
\frac{\partial P_Y(p)}{\partial\lambda}
=
-\frac{e_W(p)}{\{\lambda+Q_W(p)\}^2}
<0;
\label{eq:dPdlambda}
\end{equation}
\item[(iii)] if $\mu=0$, then $P_Y(p)=P_W(p)$ for all $p$, i.e.\ persistence is scale-free;
\item[(iv)] if $0<\xi<1$ and $Q_W(p)\to\infty$ as $p\uparrow1$, then for any fixed $\lambda$, $\lim_{p\uparrow1}P_Y(p)=\lim_{p\uparrow1}P_W(p)=1/(1-\xi)$, so the asymptotic persistence is location--scale invariant.
\end{enumerate}
\end{proposition}
The proof of Proposition~\ref{prop:locScale} is given in Appendix~\ref{app:proofLocScale}. 
Proposition~\ref{prop:locScale} is the key structural result behind the framework of Section~\ref{subsec:generalBayesian_framework}. If $Y_i=\mu_i+\sigma_iW$, $i=1,\ldots,n$, share a common shape parameter $\xi$, then (a) the asymptotic persistence $P_\infty=1/(1-\xi)$ is \emph{common} to all $i$, while (b) the finite-level persistence $P_i(p)$ varies with $i$, and does so \emph{only} through the scalar $\lambda_i=\mu_i/\sigma_i$, monotonically decreasingly. We now make $Q_W$ and $e_W$ explicit for the two families of interest.

\begin{proposition}[Persistence under the generalized Pareto distribution]
\label{prop:gpdPersistence}
Let $Z\sim GPD(\sigma,\xi)$ with $\xi<1$, and let $Y=v+Z$ for a constant $v\ge0$ (e.g.\ a threshold). For $p\in(0,1)$ write
\begin{equation}
a_p := (1-p)^{-\xi},
\qquad
Q_W(p)=\frac{a_p-1}{\xi}
\quad(\xi\neq0),
\qquad
Q_W(p)=-\log(1-p)\quad(\xi=0),
\label{eq:gpdStdQuantile}
\end{equation}
for the quantile function of the standard GPD $W\sim GPD(1,\xi)$. Then:
\begin{enumerate}
\item[(a)] the quantile mean excess of $W$ is
\begin{equation}
e_W(p)=\frac{a_p}{1-\xi}=\frac{(1-p)^{-\xi}}{1-\xi};
\label{eq:gpdMeanExcess}
\end{equation}
\item[(b)] the persistence of the \emph{excess} $Z$ does not depend on $\sigma$ and equals
\begin{equation}
P_Z(p)
=
\frac{1}{1-\xi}
+
\frac{\xi}{(1-\xi)\,(a_p-1)},
\qquad \xi\neq0,
\qquad\quad
P_Z(p)=1+\frac{1}{-\log(1-p)},
\qquad \xi=0;
\label{eq:gpdPZ}
\end{equation}
\item[(c)] the persistence of the \emph{level} $Y=v+Z$ is
\begin{equation}
P_Y(p)
=
1+
\frac{a_p}{(1-\xi)\left\{\kappa^{-1}+Q_W(p)\right\}},
\qquad
\kappa := \frac{\sigma}{v},
\label{eq:gpdPY}
\end{equation}
which depends on $(\sigma,v)$ only through $\kappa$ and is strictly increasing in $\kappa$;
\item[(d)] $\lim_{p\uparrow1}P_Z(p)=\lim_{p\uparrow1}P_Y(p)=1/(1-\xi)$ if $0<\xi<1$, and $=1$ if $\xi\le0$.
\end{enumerate}
\end{proposition}
The proofs of Propositions~\ref{prop:gpdPersistence} and~\ref{prop:gevPersistence} are given in Appendices~\ref{app:proofGPD} and~\ref{app:proofGEV}, respectively; for the GPD, part~(c) is the normalized form of the classical expected-shortfall formula \citep{McNeil2015}.

\begin{proposition}[Persistence under the generalized extreme value distribution]
\label{prop:gevPersistence}
Let $Y\sim GEV(\mu,\sigma,\xi)$ with $\xi<1$, and let $W\sim GEV(0,1,\xi)$, so that $Y=\mu+\sigma W$. For $p\in(0,1)$ write
\begin{equation}
\ell_p := -\log p,
\qquad
z_p := Q_W(p)=
\begin{cases}
\dfrac{\ell_p^{-\xi}-1}{\xi}, & \xi\ne0,\\[2mm]
-\log\ell_p, & \xi=0,
\end{cases}
\label{eq:gevStdQuantile}
\end{equation}
and let $\gamma(a,x)=\int_0^x w^{a-1}e^{-w}\,dw$ denote the lower incomplete gamma function. Then:
\begin{enumerate}
\item[(a)] the quantile mean excess of $W$ is, for $\xi\neq0$,
\begin{equation}
e_W(p)
=
\frac{1}{\xi}\left\{\frac{\gamma(1-\xi,\ell_p)}{1-p}-\ell_p^{-\xi}\right\}
\;=\;
\frac{\ell_p^{-\xi}}{\xi}\left\{G_\xi(p)-1\right\},
\qquad
G_\xi(p):=\frac{\ell_p^{\,\xi}\,\gamma(1-\xi,\ell_p)}{1-p},
\label{eq:gevMeanExcess}
\end{equation}
and, for $\xi=0$, $e_W(p)=(1-p)^{-1}\int_p^1\{-\log(-\log s)\}\,ds+\log\ell_p$; in all cases $e_W(p)\ge0$;
\item[(b)] for every $p$ with $Q_Y(p)>0$,
\begin{equation}
P_Y(p)
=
1+\frac{e_W(p)}{\lambda+z_p},
\qquad
\lambda=\frac{\mu}{\sigma},
\label{eq:gevPY}
\end{equation}
which depends on $(\mu,\sigma)$ only through $\lambda$ and is strictly decreasing in $\lambda$;
\item[(c)] for $\xi\neq0$, \eqref{eq:gevPY} admits the weighted-average form
\begin{equation}
P_Y(p)
=
\frac{c + \omega_p\,G_\xi(p)}{c+\omega_p},
\qquad
c:=\lambda-\frac{1}{\xi}=\frac{\mu-\sigma/\xi}{\sigma},
\qquad
\omega_p:=\frac{\ell_p^{-\xi}}{\xi},
\label{eq:gevWeighted}
\end{equation}
and $G_\xi(p)$ is the persistence of the GEV with $c=0$, i.e.\ $P_Y(p)=G_\xi(p)$ whenever $\mu=\sigma/\xi$;
\item[(d)] $\lim_{p\uparrow1}G_\xi(p)=\lim_{p\uparrow1}P_Y(p)=1/(1-\xi)$ if $0<\xi<1$, and $\lim_{p\uparrow1}P_Y(p)=1$ if $\xi\le0$.
\end{enumerate}
\end{proposition}

\begin{remark}[Interpretation of the location--scale ratio]
\label{rem:interpretLambda}
For $\xi>0$ the GEV has lower endpoint $\mu-\sigma/\xi$, so $c=(\mu-\sigma/\xi)/\sigma$ in~\eqref{eq:gevWeighted} is the lower endpoint expressed in units of the scale. When $c=0$ the GEV is a pure-scale Fr\'echet variable and $P_Y(p)=G_\xi(p)$, which converges to $1/(1-\xi)$ from above extremely fast (Table~\ref{tab:persistenceValues}). When $c>0$, \eqref{eq:gevWeighted} is a weighted average of $1$ and $G_\xi(p)$, so $1<P_Y(p)<G_\xi(p)$: a positive lower endpoint ``dilutes'' the escalation and persistence approaches $1/(1-\xi)$ from below. When $c<0$, $P_Y(p)>G_\xi(p)$ and persistence approaches its limit from above. Analogously, for the GPD level $Y=v+Z$, the ratio $\kappa=\sigma/v$ in~\eqref{eq:gpdPY} measures the typical exceedance size relative to the threshold, and larger $\kappa$ yields stronger finite-level persistence.
\end{remark} 
Table~\ref{tab:persistenceValues} reports exact values of \eqref{eq:gpdPZ} and \eqref{eq:gevPY} for representative shape parameters and levels. Three features are worth noting: (i) finite-level persistence is always larger than one, also for $\xi\le0$, where $P_\infty=1$; (ii) at a fixed level $p$, persistence increases with $\xi$; and (iii) for the GEV, the location--scale ratio $\lambda$ affects persistence materially at moderate levels, but its influence vanishes as $p\uparrow1$ when $\xi>0$, as predicted by Proposition~\ref{prop:locScale}(iv).

\begin{table}[ht]
\centering
\caption{Exact persistence at finite levels $p$ for the GPD excess, $P_Z(p)$ in~\eqref{eq:gpdPZ}, and for the GEV, $P_Y(p)$ in~\eqref{eq:gevPY}, at two values of the location--scale ratio $\lambda=\mu/\sigma$. The column $P_\infty$ gives the asymptotic persistence ($1/(1-\xi)$ for $\xi>0$ and $1$ for $\xi\le0$).}
\label{tab:persistenceValues}
\setlength{\tabcolsep}{4pt}
\begin{tabular}{@{}rr ccc ccc ccc@{}}
\toprule
 & & \multicolumn{3}{c}{GPD excess $P_Z(p)$} & \multicolumn{3}{c}{GEV, $\lambda=2$} & \multicolumn{3}{c}{GEV, $\lambda=5$} \\
\cmidrule(lr){3-5}\cmidrule(lr){6-8}\cmidrule(lr){9-11}
$\xi$ & $P_\infty$ & $0.9$ & $0.99$ & $0.999$ & $0.9$ & $0.99$ & $0.999$ & $0.9$ & $0.99$ & $0.999$ \\
\midrule
$-0.2$ & 1.000 & 1.285 & 1.110 & 1.056 & 1.143 & 1.066 & 1.036 & 1.080 & 1.042 & 1.024 \\
$0.0$  & 1.000 & 1.434 & 1.217 & 1.145 & 1.242 & 1.152 & 1.112 & 1.142 & 1.104 & 1.084 \\
$0.1$  & 1.111 & 1.540 & 1.301 & 1.223 & 1.316 & 1.225 & 1.186 & 1.190 & 1.163 & 1.148 \\
$0.2$  & 1.250 & 1.677 & 1.415 & 1.334 & 1.417 & 1.329 & 1.294 & 1.257 & 1.251 & 1.250 \\
$0.3$  & 1.429 & 1.859 & 1.572 & 1.490 & 1.555 & 1.478 & 1.451 & 1.352 & 1.382 & 1.403 \\
$0.4$  & 1.667 & 2.108 & 1.792 & 1.712 & 1.750 & 1.691 & 1.675 & 1.490 & 1.577 & 1.627 \\
\bottomrule
\end{tabular}
\end{table}

\subsection{A general Bayesian hierarchical framework for persistence}
\label{subsec:generalBayesian_framework}

\subsubsection{Why a constant shape parameter}
\label{subsec:whyConstantShape}

A natural first approach to nonstationary persistence is to let the shape parameter vary with time and covariates, $\xi_i=h_\xi(\mathbf x_{\xi i}^\top\boldsymbol\beta_\xi+b_{\xi i})$, so that $P_i=1/(1-\xi_i)$ varies with $i$. We do not adopt this formulation, for the following reasons.

\begin{enumerate}
\item \emph{Identifiability.} In a block maxima or threshold exceedance series, each time index $i$ typically contributes a single observation. The tail index describes the behaviour of a distribution far beyond the bulk of the data, and a single draw carries almost no information about it: the one-observation log-likelihood in $\xi_i$ is close to monotone over most of the range of $(y_i-\mu_i)/\sigma_i$, so that the posterior of $\xi_i$ is essentially determined by the prior, the link bounds and the hierarchical pooling, rather than by the data.
\item \emph{Confounding with location and scale.} With $(\mu_i,\sigma_i,\xi_i)$ all observation-specific, the data at time $i$ cannot distinguish a larger $\xi_i$ from a larger $\sigma_i$ or a shifted $\mu_i$; the resulting trade-offs produce strongly correlated, slowly mixing posteriors and, in the hierarchical setting, a feedback loop in which drift of a few $\xi_i$ toward a link boundary inflates the variance component of the $\xi$-channel, which in turn permits further drift.
\item \emph{Interpretation of persistence.} Theorem~\ref{thm:persistence} is a statement about the tail of a single distribution, and $P_\infty=1/(1-\xi)$ is meaningful only as a property of that distribution. Under a time-varying $\xi_i$ informed by one observation, a posterior trajectory of $1/(1-\xi_i)$ would largely reflect prior structure rather than a data-supported change in tail heaviness.
\item \emph{Standard practice.} In regression models for extremes the shape parameter is usually held constant while location and scale vary, precisely because it is the most difficult parameter to estimate \citep{Coles2001}.
\end{enumerate}
An important thing to note is that holding $\xi$ constant does \emph{not} remove nonstationarity from persistence. By Proposition~\ref{prop:locScale}, a common $\xi$ together with time-varying $(\mu_i,\sigma_i)$ yields (a) a single, well-identified asymptotic persistence $P_\infty=1/(1-\xi)$, pooled over all $n$ observations, and (b) a time-varying finite-level persistence profile $P_i(p)$, driven by the location--scale ratio $\lambda_i=\mu_i/\sigma_i$, which is identified through the location and scale channels, both of which are informed by the bulk of the extremal data. 

\subsubsection{Bayesian hierarchical persistence modeling framework}
\label{subsec:hierarchicalModel}

Let $Y_1,\ldots,Y_n$ denote independent observations arising from an extreme value model with parameter vector
$\Theta_i=(\theta_{1i},\ldots,\theta_{Ki},\xi)$, $i=1,\ldots,n$, where
$\xi$ denotes the extreme value index, \emph{common to all observations}, and $\theta_{1i},\ldots,\theta_{Ki}$
represent the remaining, observation-specific distributional parameters (e.g., location and
scale), with $f(\cdot)$ left unspecified so that the framework accommodates
the GEV, generalized Pareto, or other extreme value families
interchangeably. Rather than linking each $\theta_{ji}$ directly to covariates, it is convenient to introduce, for every non-tail parameter channel
$j=1,\ldots,K$, an
\emph{unconstrained} latent linear predictor $\widetilde\theta_{ji}\in\mathbb{R}$
together with a known bijection $h_j:\mathbb{R}\to\Xi_j$ onto the natural
parameter space $\Xi_j$ of $\theta_j$, such that
\begin{equation}
\theta_{ji} = h_j\!\left(\widetilde\theta_{ji}\right),
\qquad
\widetilde\theta_{ji} = \mathbf{x}_{ji}^{\top}\boldsymbol{\beta}_j + b_{ji},
\qquad
j=1,\ldots,K,
\label{eq:tildeLinkGeneral}
\end{equation}
and, for the tail index, a \emph{single} unconstrained scalar $\widetilde\xi\in\mathbb{R}$ with
\begin{equation}
\xi = h_\xi\!\left(\widetilde\xi\right),
\qquad
\widetilde\xi \sim N\!\left(m_\xi,\,s_\xi^2\right),
\label{eq:tildeLinkXiGeneral}
\end{equation}
where $m_\xi$ and $s_\xi^2$ are fixed hyperparameters. We assume that $b_{ji}\sim N(0,\tau_j^2)$, and hence \eqref{eq:tildeLinkGeneral}
is equivalent to the compact conditional statement
\begin{equation}
\widetilde\theta_{ji} \mid \boldsymbol{\beta}_j,\tau_j^2
\;\sim\;
N\!\left(\mathbf{x}_{ji}^{\top}\boldsymbol{\beta}_j,\,\tau_j^2\right),
\qquad
j=1,\ldots,K,\quad i=1,\ldots,n.
\label{eq:tildeConditional}
\end{equation}
Writing the model in terms of $h_j$ and the tilde notation makes explicit that each
latent linear predictor, and the tail-index predictor $\widetilde\xi$, is exactly Gaussian a priori -- a property
exploited directly in posterior computation
(Section~\ref{subsec:generalMHGsteps}; the full conditionals are derived in Section~\xref{S-sec:supp_mcmc} of the Supplementary Material). The link function
$g_j(\cdot) := h_j^{-1}(\cdot)$ recovers the usual notation
$g_j(\theta_{ji}) = \mathbf{x}_{ji}^{\top}\boldsymbol{\beta}_j+b_{ji}$; the
two are used interchangeably below, with $h_j$ preferred when emphasizing
the forward mapping from the unconstrained scale to $\theta_j$'s natural
parameter space, and $g_j$ preferred when emphasizing the reverse
(link) direction. The full hierarchical specification is given jointly by
\begin{equation}
\begin{aligned}
Y_i \mid \Theta_i &\sim f(y_i \mid \theta_{1i},\ldots,\theta_{Ki},\xi),
&& i=1,\ldots,n, \\[4pt]
\theta_{ji} &= h_j\!\left(\widetilde\theta_{ji}\right),
\qquad
\widetilde\theta_{ji} = \mathbf{x}_{ji}^{\top}\boldsymbol{\beta}_j + b_{ji},
&& j=1,\ldots,K, \\[4pt]
\xi &= h_\xi\!\left(\widetilde\xi\right),
\qquad
\widetilde\xi \sim N\!\left(m_\xi,\,s_\xi^2\right),
&& \\[4pt]
\boldsymbol{\beta}_j &\sim N\!\left(\mathbf{0},\,\sigma_{\beta_j}^2\,I_{p_j}\right),
&& j=1,\ldots,K, \\[4pt]
b_{ji} &\sim N(0,\tau_j^2),
\qquad
\tau_j^2 \sim IG(a_j,b_j),
&& j=1,\ldots,K,\ i=1,\ldots,n,
\end{aligned}
\label{eq:fullHierarchy}
\end{equation}
where $\mathbf{x}_{ji}$ is the covariate vector for channel $j$,
$\boldsymbol{\beta}_j\in\mathbb{R}^{p_j}$ is the associated regression
coefficient vector, $b_{ji}$ is an optional latent effect accounting for
unexplained heterogeneity or temporal dependence, and $\sigma_{\beta_j}^2$,
$(a_j,b_j)$, $(m_\xi,s_\xi^2)$ are fixed hyperparameters. In particular, $\sigma_{\beta_j}^2$
is treated as a fixed design constant (e.g.\ $\sigma_{\beta_j}^2 = 10$ for
all $j$, yielding a weakly informative prior) rather than assigned its own
hyperprior, so that only the latent-effect variances $\tau_j^2$ are
themselves random and require an $IG(a_j,b_j)$ hyperprior. Compared with a channel-wise regression for $\xi$, the tail-index channel in~\eqref{eq:fullHierarchy} carries no regression coefficients, latent effects or variance component: it reduces to a single scalar parameter, informed jointly by all $n$ observations. Alternative
prior specifications for $\boldsymbol\beta_j$, $\tau_j^2$ and $\widetilde\xi$ may be adopted
depending on the application without altering the proposed framework.

\paragraph{Induced prior on $\xi$ and on $P_\infty$.}
Since $h_\xi$ is a strictly increasing bijection, the prior \eqref{eq:tildeLinkXiGeneral} induces on $\xi$ the density
\begin{equation}
\pi(\xi)
=
\phi\!\left(\frac{h_\xi^{-1}(\xi)-m_\xi}{s_\xi}\right)
\frac{1}{s_\xi}
\left|\frac{d\,h_\xi^{-1}(\xi)}{d\xi}\right|,
\qquad \xi\in h_\xi(\mathbb{R}),
\label{eq:inducedXiPrior}
\end{equation}
where $\phi$ denotes the standard normal density. For the bounded logistic map adopted in Section~\ref{subsec:illust_example},
$h_\xi(w)=\xi_{\min}+(\xi_{\max}-\xi_{\min})\,\mathrm{expit}(w)$, one has
$h_\xi^{-1}(\xi)=\operatorname{logit}\{(\xi-\xi_{\min})/(\xi_{\max}-\xi_{\min})\}$ and
\[
\frac{d\,h_\xi^{-1}(\xi)}{d\xi}=\frac{\xi_{\max}-\xi_{\min}}{(\xi-\xi_{\min})(\xi_{\max}-\xi)}.
\]
Choosing $m_\xi=0$ places the prior median of $\xi$ at the midpoint $(\xi_{\min}+\xi_{\max})/2$, and $s_\xi\approx1.7$ yields an induced prior on $\xi$ close to uniform on $(\xi_{\min},\xi_{\max})$ (exact uniformity would correspond to a standard logistic rather than a normal prior on $\widetilde\xi$). The prior on $P_\infty$ follows by the further monotone change of variables $P_\infty=1/(1-\xi)$, $d\xi/dP_\infty=P_\infty^{-2}$.

\begin{remark}[Boundedness of $h_\xi$ and persistence]
\label{rem:generalBoundedLink}
Because $P_\infty = 1/(1-\xi)$ by Theorem~\ref{thm:persistence}, the asymptotic
persistence is unbounded whenever $\xi$ can approach $1$ with positive
posterior probability. If $h_\xi(\cdot)$ in~\eqref{eq:tildeLinkXiGeneral}
is chosen to map $\mathbb{R}$ onto a bounded interval
$\Xi_\xi = (\xi_{\min},\xi_{\max}) \subset (-\infty,1)$ -- as opposed to an
unrestricted or half-bounded range -- then $P_\infty \le 1/(1-\xi_{\max})$ holds
almost surely under the prior and, consequently, under the posterior. The bound $\xi_{\max}<1$ also guarantees that the quantile integrals defining the finite-level persistence in Propositions~\ref{prop:gpdPersistence}--\ref{prop:gevPersistence} are finite. This
is a property of the \emph{link} $h_\xi$, not of the extreme value family
$f(\cdot)$ itself, and therefore applies uniformly across GEV, GPD, or
other specializations of the framework.
\end{remark}

\subsubsection{Posterior computation}
\label{subsec:generalMHGsteps}

Posterior inference under~\eqref{eq:fullHierarchy} is carried out by a Metropolis-within-Gibbs sampler in which the latent predictors $\widetilde\theta_{ji}$ and the scalar $\widetilde\xi$, rather than standardized latent effects, are stored between iterations. Under this parameterization the full conditional of every $\boldsymbol\beta_j$ is Gaussian and that of every $\tau_j^2$ is inverse-gamma, so all regression coefficients and variance components are updated by direct Gibbs draws. Each $\widetilde\theta_{ji}$ is updated by random-walk Metropolis using only the $i$th likelihood term, with proposal scales adapted during burn-in only. The common predictor $\widetilde\xi$, whose full conditional involves the entire likelihood, is updated by elliptical slice sampling (ESS) \citep{murray2010elliptical} applied to $\eta=(\widetilde\xi-m_\xi)/s_\xi$, which has an exact $N(0,1)$ prior. This update requires no tuning, always moves, and enforces the support constraints of $f$ automatically; it is preferable to a random walk because the bounded map $h_\xi$ flattens the likelihood near $\xi_{\min}$ and $\xi_{\max}$. At iteration $m$, the shape and the asymptotic persistence are recovered as
\begin{equation}
\widetilde\xi^{(m)} = m_\xi + s_\xi\eta^{(m)},
\qquad
\xi^{(m)} = h_\xi\!\left(\widetilde\xi^{(m)}\right),
\qquad
P_\infty^{(m)}=\frac{1}{1-\xi^{(m)}}\;\le\;\frac{1}{1-\xi_{\max}},
\label{eq:xiRecover}
\end{equation}
the inequality being the guarantee of Remark~\ref{rem:generalBoundedLink}, realized draw by draw. The joint posterior, the full conditional distributions of every block, the complete sampler (Algorithm~\xref{S-alg:MHG}) and the additional interweaving, PIT-preserving and joint moves used in Sections~\ref{sec:simulation}--\ref{sec:application} are given in Section~\xref{S-sec:supp_mcmc} of the Supplementary Material.

\subsubsection{Posterior inference of persistence}
\label{subsec:posteriorPersistenceInference}

The sampler of Section~\ref{subsec:generalMHGsteps} (Algorithm~\xref{S-alg:MHG} in the Supplementary Material) produces a sequence of stored states
$\{\Psi^{(m)}\}_{m=1}^{N}$, where
$\Psi^{(m)} = \big(\widetilde\Theta^{(m)},\boldsymbol\beta^{(m)},\boldsymbol\tau^{2(m)}\big)$
collects the latent predictors $\widetilde\Theta$ (including the scalar $\widetilde\xi$), regression
coefficients $\boldsymbol\beta$, and variance components $\boldsymbol\tau^2$; no separate latent-effect vector $\mathbf b$ need be stored, since
$b_{ji}^{(m)} = \widetilde\theta_{ji}^{(m)} - \mathbf x_{ji}^\top\boldsymbol\beta_j^{(m)}$
is recovered from $\Psi^{(m)}$ whenever required. The post-processing step
retains every $L$th draw after discarding the first $B$ as burn-in,
yielding $M = \lfloor (N-B)/L \rfloor$ retained samples, with each $\Theta^{(m)}$ obtained from
$\widetilde\Theta^{(m)}$ via the channel maps $h_j$, $j=1,\ldots,K$, and $h_\xi$. Two persistence summaries are computed from each retained draw.

\paragraph{Asymptotic persistence.} The draws $\{P_\infty^{(m)}\}_{m=1}^M$ of~\eqref{eq:xiRecover} provide the posterior of the single, time-invariant quantity $P_\infty=1/(1-\xi)$: its posterior mean $\widehat P_\infty$, equal-tailed credible interval from the empirical quantiles of $\{P_\infty^{(m)}\}$, and exceedance probabilities $\Pr(P_\infty>P_0\mid\mathbf y)$. Since $P_\infty$ carries its regular-variation meaning only for $\xi>0$ (Remark~\ref{rem:regularVariationCaveat}), we always report
\begin{equation}
\Pr(\xi>0\mid\mathbf y)\approx\frac{1}{M}\sum_{m=1}^M\mathbf 1\{\xi^{(m)}>0\}
\label{eq:probXiPos}
\end{equation}
alongside it. By Remark~\ref{rem:generalBoundedLink},
\begin{equation}
P_\infty^{(m)} \;\le\; \frac{1}{1-\xi_{\max}},
\qquad
m=1,\ldots,M,
\label{eq:persistenceBoundRetained}
\end{equation}
so the reported posterior summaries are guaranteed finite and stable.

\paragraph{Finite-level, time-varying persistence.} For a chosen level $p\in[p_0,1)$ (or a grid of levels), the finite-level persistence of observation $i$ is obtained by plugging each retained draw into the closed forms of Propositions~\ref{prop:gpdPersistence}--\ref{prop:gevPersistence}:
\begin{equation}
P_i^{(m)}(p)
=
1+\frac{e_W\!\left(p;\xi^{(m)}\right)}{\lambda_i^{(m)}+Q_W\!\left(p;\xi^{(m)}\right)},
\qquad
m=1,\ldots,M,\ i=1,\ldots,n,
\label{eq:posteriorPersistenceDraw}
\end{equation}
where $\lambda_i^{(m)}$ is the model-specific location--scale ratio (Section~\ref{subsec:illust_example}) evaluated at $\Theta_i^{(m)}$. These draws yield pointwise posterior means $\widehat P_i(p)$, credible bands for the persistence trajectory $\{P_i(p)\}_{i=1}^n$, and exceedance probabilities $\Pr(P_i(p)>P_0\mid\mathbf y)$. Moreover, since $\xi$ is common to $i$ and $i'$ and \eqref{eq:PlocScale} is strictly decreasing in $\lambda$, comparisons of persistence across time reduce exactly to comparisons of the location--scale ratio,
\begin{equation}
\Pr\!\left(P_i(p)>P_{i'}(p)\,\middle|\,\mathbf y\right)
=
\Pr\!\left(\lambda_i<\lambda_{i'}\,\middle|\,\mathbf y\right),
\qquad\text{for every } p,
\label{eq:persistenceContrast}
\end{equation}
so that the posterior probability of an increase in persistence between two time points does not depend on the level $p$ at which it is assessed (Proposition~\ref{prop:contrast} in Appendix~\ref{app:proofContrast}). No additional post-hoc computation beyond thinning and evaluation of the closed forms is required.

\subsection{Illustrative extreme value models}
\label{subsec:illust_example}

The Bayesian persistence framework developed above is applicable to any
extreme value model possessing a well-defined tail index. To illustrate its
implementation, we consider the two principal models in univariate extreme
value analysis, namely the generalized Pareto distribution for threshold
exceedances and the generalized extreme value distribution for block
maxima. In both cases the shape parameter is common across observations and linked through the bounded map
\begin{equation}
\xi=h_\xi(\widetilde\xi)=\xi_{\min}+(\xi_{\max}-\xi_{\min})\,\mathrm{expit}(\widetilde\xi),
\qquad
\widetilde\xi\sim N(m_\xi,s_\xi^2),
\qquad
-\infty<\xi_{\min}<\xi_{\max}<1,
\label{eq:boundedXiLink}
\end{equation}
with fixed bounds chosen a priori, so that the resulting asymptotic persistence is guaranteed finite by construction. Allowing $\xi_{\min}\le0$ lets the posterior place mass in the Weibull ($\xi<0$), Gumbel ($\xi=0$), and Fr\'echet ($\xi>0$) domains
within a single fit. Both models are written in the latent-linear-predictor form
\eqref{eq:tildeLinkGeneral}--\eqref{eq:tildeLinkXiGeneral}, so that every regression coefficient
$\boldsymbol\beta_j$ and every latent-effect variance $\tau_j^2$
is updated by a direct conjugate Gibbs draw, the observation-specific predictors $\widetilde\theta_{ji}$ by adaptive random-walk Metropolis, and the scalar $\widetilde\xi$ by elliptical slice sampling.

\subsubsection{Generalized Pareto Model}
\label{subsec:gpd_model}

Suppose that $Y_1,\ldots,Y_n$ denote observations exceeding
sufficiently high (possibly covariate-dependent) thresholds $v_i$, and let $Z_i=Y_i-v_i$ denote the excesses. Under the peak-over-threshold framework,
\[
Z_i
\sim
GPD(\sigma_i,\xi),
\qquad i=1,\ldots,n,
\]
where $\sigma_i>0$ is the observation-specific scale and $\xi$ the common shape parameter (see Section~\ref{subsec:pot_details}). When
viewed through the general framework of
Section~\ref{subsec:generalBayesian_framework}, here $K=1$, with
$\theta_{1i}=\sigma_i$. To accommodate
nonstationarity, the scale parameter is linked to explanatory variables
through the log map, $h_\sigma = \exp(\cdot)$,
\begin{equation}
\widetilde\sigma_i
=
\mathbf{x}_{\sigma i}^{\top}\boldsymbol{\beta}_{\sigma}
+b_{\sigma i},
\qquad
\sigma_i = \exp\!\left(\widetilde\sigma_i\right),
\qquad
\widetilde\sigma_i \mid \boldsymbol\beta_\sigma,\tau_\sigma^2
\;\sim\;
N\!\left(\mathbf{x}_{\sigma i}^{\top}\boldsymbol{\beta}_{\sigma},\,\tau_\sigma^2\right),
\label{eq:gpdSigmaLink}
\end{equation}
while the tail index is common and given by~\eqref{eq:boundedXiLink}, with
$\boldsymbol\beta_\sigma \sim N(\mathbf 0,\sigma_{\beta_\sigma}^2 I)$ and $\tau_\sigma^2\sim IG(a_\sigma,b_\sigma)$, per~\eqref{eq:fullHierarchy}.

\paragraph{Likelihood.} Writing $t_i:=1+\xi z_i/\sigma_i$, the log-likelihood is
\begin{equation}
\ell_{\mathrm{GPD}}
=
\sum_{i=1}^n\left\{-\log\sigma_i-\left(1+\frac1\xi\right)\log t_i\right\},
\qquad t_i>0\ \ \forall i,
\label{eq:gpdLogLik}
\end{equation}
($\xi\ne0$; for $\xi=0$ the summand is $-\log\sigma_i-z_i/\sigma_i$). For $\xi<0$, the support constraint $t_i>0$ requires $\xi>\max_i(-\sigma_i/z_i)$; in the sampler it is enforced by setting $\ell=-\infty$. The update of each $\widetilde\sigma_i$ involves only the $i$th summand of~\eqref{eq:gpdLogLik}, whereas the update of $\widetilde\xi$ involves the full sum; the full conditionals are given in Section~\xref{S-sec:supp_mcmc_gpd} of the Supplementary Material.

\paragraph{Persistence under the GPD model.} Two persistence summaries are relevant.
\begin{enumerate}
\item \emph{Asymptotic persistence.} At each iteration,
\begin{equation}
\xi^{(m)}=h_\xi\!\left(\widetilde\xi^{(m)}\right),
\qquad
P_\infty^{(m)}
=
\frac{1}{1-\xi^{(m)}}
\;\le\;
\frac{1}{1-\xi_{\max}},
\label{eq:gpdPersistence}
\end{equation}
giving posterior inference for the time-invariant asymptotic persistence of the exceedance tail, accompanied by $\Pr(\xi>0\mid\mathbf y)$.

\item \emph{Finite-level persistence of the excesses.} By Proposition~\ref{prop:gpdPersistence}(b), the persistence of the excess $Z_i$ at level $p$,
\begin{equation}
P_{Z}^{(m)}(p)
=
\frac{1}{1-\xi^{(m)}}+\frac{\xi^{(m)}}{\left(1-\xi^{(m)}\right)\left\{(1-p)^{-\xi^{(m)}}-1\right\}},
\label{eq:gpdPZdraw}
\end{equation}
does \emph{not} depend on $\sigma_i$, and is therefore the same for every $i$. Under a common shape parameter, the GPD excess persistence is time-invariant at \emph{every} level $p$, not only asymptotically: nonstationarity in $\sigma_i$ changes the absolute size of the excesses, but not their relative escalation beyond a quantile of the excess distribution.

\item \emph{Finite-level persistence of the levels.} When interest lies in the escalation of the observed variable itself, the relevant quantity is the persistence of $Y_i=v_i+Z_i$ given an exceedance, which by Proposition~\ref{prop:gpdPersistence}(c) is
\begin{equation}
P_{Y,i}^{(m)}(p)
=
1+\frac{(1-p)^{-\xi^{(m)}}}{\left(1-\xi^{(m)}\right)\left\{\lambda_i^{(m)}+Q_W\!\left(p;\xi^{(m)}\right)\right\}},
\qquad
\lambda_i^{(m)}=\frac{v_i}{\sigma_i^{(m)}},
\label{eq:gpdPYdraw}
\end{equation}
with $Q_W(p;\xi)=\{(1-p)^{-\xi}-1\}/\xi$. This varies with $i$ through the threshold-to-scale ratio $\lambda_i=v_i/\sigma_i$: persistence of the levels increases over time exactly when the typical exceedance size $\sigma_i$ grows faster than the threshold $v_i$, and \eqref{eq:persistenceContrast} gives $\Pr(P_{Y,i}(p)>P_{Y,i'}(p)\mid\mathbf y)=\Pr(v_i/\sigma_i<v_{i'}/\sigma_{i'}\mid\mathbf y)$. If $v_i\equiv v$ is constant, this reduces to $\Pr(\sigma_i>\sigma_{i'}\mid\mathbf y)$.
\end{enumerate}
If, in addition, the exceedance probability $\zeta_i=\Pr(Y_i>v_i)$ is modelled, persistence of the parent distribution at an unconditional level $q\ge1-\zeta_i$ follows from the same expressions with $p=1-(1-q)/\zeta_i$, since the conditional tail beyond the $q$th parent quantile coincides with the conditional tail of the excess distribution beyond its $p$th quantile (Proposition~\ref{prop:uncond}).

\vspace{2mm}
\subsubsection{Generalized Extreme Value Model}
\label{subsec:gev_model}

For block maxima, let
\[
Y_i
\sim
GEV(\mu_i,\sigma_i,\xi),
\qquad i=1,\ldots,n,
\]
where $\mu_i$ and $\sigma_i$ denote the observation-specific location and scale and $\xi$ the common shape parameter. Here $K=2$, with
$(\theta_{1i},\theta_{2i})=(\mu_i,\sigma_i)$. A flexible regression
specification is obtained through the identity map $h_\mu=\mathrm{id}(\cdot)$
for location and the log map $h_\sigma=\exp(\cdot)$ for scale,
\begin{align}
\widetilde\mu_i
&=
\mathbf{x}_{\mu i}^{\top}\boldsymbol{\beta}_{\mu}
+b_{\mu i},
&
\mu_i &= \widetilde\mu_i,
&
\widetilde\mu_i \mid \boldsymbol\beta_\mu,\tau_\mu^2
&\sim
N\!\left(\mathbf{x}_{\mu i}^{\top}\boldsymbol{\beta}_{\mu},\,\tau_\mu^2\right),
\label{eq:gevMuLinkIllust}
\\
\widetilde\sigma_i
&=
\mathbf{x}_{\sigma i}^{\top}\boldsymbol{\beta}_{\sigma}
+b_{\sigma i},
&
\sigma_i &= \exp\!\left(\widetilde\sigma_i\right),
&
\widetilde\sigma_i \mid \boldsymbol\beta_\sigma,\tau_\sigma^2
&\sim
N\!\left(\mathbf{x}_{\sigma i}^{\top}\boldsymbol{\beta}_{\sigma},\,\tau_\sigma^2\right),
\label{eq:gevSigmaLinkIllust}
\end{align}
and the common tail index~\eqref{eq:boundedXiLink}, with $\boldsymbol\beta_\mu\sim N(\mathbf 0,\sigma_{\beta_\mu}^2 I)$,
$\boldsymbol\beta_\sigma\sim N(\mathbf 0,\sigma_{\beta_\sigma}^2 I)$ and
$\tau_\mu^2,\tau_\sigma^2 \sim IG(a_j,b_j)$ as
in~\eqref{eq:fullHierarchy}; suitable covariates may be
incorporated into either or both of the location and scale channels.

\paragraph{Likelihood.} Writing $t_i:=1+\xi(y_i-\mu_i)/\sigma_i$, the log-likelihood is
\begin{equation}
\ell_{\mathrm{GEV}}
=
\sum_{i=1}^n\left\{-\log\sigma_i-\left(1+\frac1\xi\right)\log t_i-t_i^{-1/\xi}\right\},
\qquad t_i>0\ \ \forall i,
\label{eq:gevLogLik}
\end{equation}
($\xi\ne0$; for $\xi=0$ the summand is $-\log\sigma_i-r_i-e^{-r_i}$ with $r_i=(y_i-\mu_i)/\sigma_i$). The support constraint $t_i>0$ now couples $\xi$ to every $(\mu_i,\sigma_i)$: for $\xi>0$ it requires $\xi<\min_{i:\,y_i<\mu_i}\{\sigma_i/(\mu_i-y_i)\}$, and for $\xi<0$ it requires $\xi>\max_{i:\,y_i>\mu_i}\{-\sigma_i/(y_i-\mu_i)\}$. These constraints are enforced in the sampler by setting $\ell=-\infty$, both in the updates of $(\widetilde\mu_i,\widetilde\sigma_i)$, which involve only the $i$th summand of~\eqref{eq:gevLogLik}, and in the update of $\widetilde\xi$, which involves the full sum; the full conditionals are given in Section~\xref{S-sec:supp_mcmc_gev} of the Supplementary Material.

\paragraph{Persistence under the GEV model.}
\begin{enumerate}
\item \emph{Asymptotic persistence.} As in~\eqref{eq:gpdPersistence},
\begin{equation}
P_\infty^{(m)}
=
\frac{1}
{1-\xi^{(m)}}
\;\le\;
\frac{1}{1-\xi_{\max}},
\label{eq:gevPersistenceIllust}
\end{equation}
a single, time-invariant quantity describing the escalation of block maxima in the far tail, reported together with $\Pr(\xi>0\mid\mathbf y)$. By Proposition~\ref{prop:locScale}(iv), it is unaffected by the trends in $\mu_i$ and $\sigma_i$.

\item \emph{Finite-level, time-varying persistence.} By Proposition~\ref{prop:gevPersistence}, at level $p$,
\begin{equation}
P_i^{(m)}(p)
=
1+\frac{e_W\!\left(p;\xi^{(m)}\right)}{\lambda_i^{(m)}+z_p\!\left(\xi^{(m)}\right)},
\qquad
\lambda_i^{(m)}=\frac{\mu_i^{(m)}}{\sigma_i^{(m)}},
\label{eq:gevPYdraw}
\end{equation}
with $z_p(\xi)$ and $e_W(p;\xi)$ as in~\eqref{eq:gevStdQuantile}--\eqref{eq:gevMeanExcess}; $e_W$ is evaluated through the regularized incomplete gamma function, $\gamma(1-\xi,\ell_p)=\Gamma(1-\xi)\,\mathrm{P}(1-\xi,\ell_p)$, available in standard software (e.g.\ \texttt{pgamma} in R). The persistence of block maxima at a fixed return level therefore evolves over time solely through the location--scale ratio $\lambda_i=\mu_i/\sigma_i$: it increases when the scale grows relative to the location (greater relative spread of the maxima), and decreases when the location grows relative to the scale (maxima shift upward without proportionate spreading). Equation~\eqref{eq:persistenceContrast} gives the posterior probability of an increase between times $i'$ and $i$ as $\Pr(\mu_i/\sigma_i<\mu_{i'}/\sigma_{i'}\mid\mathbf y)$, at every level $p$.

\item \emph{Lower-endpoint interpretation.} For $\xi>0$, Remark~\ref{rem:interpretLambda} gives the equivalent reading through $c_i=\lambda_i-1/\xi$, the lower endpoint of the GEV in units of its scale: $c_i>0$ pulls $P_i(p)$ below the pure Fr\'echet value $G_\xi(p)$ and toward one, $c_i<0$ pushes it above, and $c_i=0$ gives $P_i(p)=G_\xi(p)\approx1/(1-\xi)$ at all high levels.
\end{enumerate}
The ratio in \eqref{eq:gevPYdraw} requires $Q_{Y,i}(p)=\mu_i+\sigma_i z_p>0$, which holds at high levels for positive-valued responses such as rainfall maxima; for responses that may be negative (e.g.\ temperature anomalies), persistence should be defined after a fixed, scientifically meaningful shift of origin, since by Lemma~\ref{lem:meanExcess} persistence is not shift-invariant.

These examples illustrate that the proposed Bayesian framework is
independent of the particular extreme value model and may be readily
adapted to both threshold exceedance and block maxima analyses. Because both models share the same bounded map $h_\xi$ for the common $\xi$, the same ESS update for $\widetilde\xi$, and the same conjugate Gibbs updates for every $\boldsymbol\beta_j$ and $\tau_j^2$, Algorithm~\xref{S-alg:MHG} applies verbatim to
either specialization, with only the likelihood $f(\cdot)$, the number of
non-tail-index channels $K$, the channel maps $h_j$, and the location--scale ratio $\lambda_i$ entering~\eqref{eq:posteriorPersistenceDraw}
changing between the two cases: $K=1$, $h_\sigma=\exp(\cdot)$ and $\lambda_i=v_i/\sigma_i$
for the generalized Pareto model, and $K=2$,
$(h_\mu,h_\sigma)=(\mathrm{id}(\cdot),\exp(\cdot))$ and $\lambda_i=\mu_i/\sigma_i$
for the generalized extreme value model. Table~\ref{tab:persistenceSummary} summarizes the resulting persistence quantities.

\begin{table}[ht]
\centering
\caption{Persistence quantities under the constant-shape GPD and GEV models. $Q_W$ and $e_W$ denote the standardized quantile and quantile mean excess of Propositions~\ref{prop:gpdPersistence}--\ref{prop:gevPersistence}.}
\label{tab:persistenceSummary}
\renewcommand{\arraystretch}{1.3}
\begin{tabularx}{\textwidth}{@{}l X X@{}}
\toprule
 & GPD ($Z_i=Y_i-v_i\sim GPD(\sigma_i,\xi)$) & GEV ($Y_i\sim GEV(\mu_i,\sigma_i,\xi)$) \\
\midrule
Asymptotic $P_\infty$ & $1/(1-\xi)$; common to all $i$ & $1/(1-\xi)$; common to all $i$ \\
Location--scale ratio & $\lambda_i=v_i/\sigma_i$ & $\lambda_i=\mu_i/\sigma_i$ \\
$Q_W(p)$ & $\{(1-p)^{-\xi}-1\}/\xi$ & $\{(-\log p)^{-\xi}-1\}/\xi$ \\
$e_W(p)$ & $(1-p)^{-\xi}/(1-\xi)$ & $\xi^{-1}\{\gamma(1-\xi,-\log p)/(1-p)-(-\log p)^{-\xi}\}$ \\
Finite-level $P_i(p)$ & excess: $P_Z(p)$ in~\eqref{eq:gpdPZ}, common to all $i$; level: $1+e_W(p)/\{\lambda_i+Q_W(p)\}$ & $1+e_W(p)/\{\lambda_i+Q_W(p)\}$ \\
Source of temporal change & $\sigma_i$ relative to $v_i$ & $\sigma_i$ relative to $\mu_i$ \\
\bottomrule
\end{tabularx}
\end{table}

\begin{remark}[Regime-wise shape parameters]
\label{rem:regimeShape}
If a change in tail heaviness itself is of scientific interest, the framework extends directly to a small number of regime-specific shape parameters, $\xi_{r}=h_\xi(\widetilde\xi_r)$, $r=1,\ldots,R$, each shared by all observations in a pre-specified epoch (e.g.\ before and after a known change point) and each informed by many observations. Each $\widetilde\xi_r$ is then updated by a separate scalar ESS step using only the likelihood terms of its regime, and $\Pr(\xi_2>\xi_1\mid\mathbf y)=\Pr(P_{\infty,2}>P_{\infty,1}\mid\mathbf y)$ provides a direct posterior assessment of a change in asymptotic persistence.
\end{remark}

\section{Simulation study}
\label{sec:simulation}

The simulation study addresses two questions: whether the parameters that determine persistence, above all the common shape $\xi$, can be recovered under the hierarchical GEV and GPD specifications of Section~\ref{subsec:illust_example}; and whether the posterior summaries of Section~\ref{subsec:posteriorPersistenceInference}, namely the asymptotic persistence $P_\infty$, the finite-level trajectories $P_i(p)$ and the ordering probabilities~\eqref{eq:persistenceContrast}, are accurate and well calibrated.

\subsection{Simulation set-up}
\label{subsec:sim-design}

\paragraph{Covariates.} 
Both models share the design vector $\mathbf x_i=(1,\,s_i,\,x_{1i},\,x_{2i})^\top$, $i=1,\ldots,n$, where $s_i$ is the standardized time index, $x_{1i}$ is a Gaussian AR(1) series with autocorrelation $0.3$ that plays the role of a climate index, and $x_{2i}$ is an AR(1) series with autocorrelation $0.1$ that has no effect in either channel. The covariates are standardized and held fixed across replicates. We consider $n\in\{100,500\}$; the smaller value is of the order of the application in Section~\ref{sec:application}.

\paragraph{GEV model.} Block maxima are generated as $Y_i\sim\mathrm{GEV}(\mu_i,\sigma_i,\xi)$, with
$\mu_i=\mathbf x_i^\top\boldsymbol\beta_\mu+b_{\mu i}$ and $\log\sigma_i=\mathbf x_i^\top\boldsymbol\beta_\sigma+b_{\sigma i}$, where $b_{ji}\overset{\mathrm{iid}}{\sim}N(0,\tau_j^2)$, and
\[
\boldsymbol\beta_\mu=(40,\,2,\,4,\,0)^\top,\quad
\boldsymbol\beta_\sigma=(\log 10,\,0.15,\,0.10,\,0)^\top,\quad
\tau_\mu^2=1,\quad \tau_\sigma^2=0.04,\quad \xi=0.20 .
\]
Hence $P_\infty=1.25$. The scale grows faster over time than the location, so the location--scale ratio $\lambda_i=\mu_i/\sigma_i$ decreases on average and, by Proposition~\ref{prop:gevPersistence}(b), the finite-level persistence increases over time.

\paragraph{GPD model.} Threshold excesses are generated as $Z_i\sim\mathrm{GPD}(\sigma_i,\xi)$ above a constant threshold $v=10$, with $\log\sigma_i=\mathbf x_i^\top\boldsymbol\beta_\sigma+b_{\sigma i}$, $b_{\sigma i}\overset{\mathrm{iid}}{\sim}N(0,\tau_\sigma^2)$,
\[
\boldsymbol\beta_\sigma=(0.50,\,0.20,\,0.15,\,0)^\top,\qquad \tau_\sigma^2=0.09,\qquad \xi=0.25,
\]
so that $P_\infty=4/3$ and, by~\eqref{eq:gpdPZ}, $P_Z(0.99)=1.487$ for every $i$. The level persistence $P_{Y,i}(p)$ of~\eqref{eq:gpdPYdraw} varies with $i$ through $\lambda_i=v/\sigma_i$.

\paragraph{Priors and computation.} For the GEV model, $\boldsymbol\beta_\mu\sim N(\mathbf 0,10^4I)$, $\boldsymbol\beta_\sigma\sim N(\mathbf 0,10I)$, $\tau_\mu^2\sim IG(2,1)$, $\tau_\sigma^2\sim IG(2,0.05)$ and $(\xi_{\min},\xi_{\max})=(0,0.8)$. For the GPD model, $\boldsymbol\beta_\sigma\sim N(\mathbf 0,100I)$, $\tau_\sigma^2\sim IG(3,0.2)$ and $(\xi_{\min},\xi_{\max})=(-0.5,0.8)$, so that the posterior can place mass in the Weibull domain and $\Pr(\xi>0\mid\mathbf y)$ is informative. In both cases $\widetilde\xi\sim N(0,1.7^2)$, which induces an approximately uniform prior on $(\xi_{\min},\xi_{\max})$ by~\eqref{eq:inducedXiPrior}. Posterior samples are drawn with Algorithm~\xref{S-alg:MHG}, augmented by the interweaving, PIT-preserving (ridge) and joint moves described in Section~\xref{S-sec:supp_moves} of the Supplementary Material. For each of $R=50$ replicates and each $(\text{model},n)$ combination, two chains are run from dispersed starting values for $8{,}000$ iterations, of which the first $3{,}000$ are discarded as burn-in, with every fifth draw retained ($2{,}000$ draws in total). For one representative data set with $n=500$, four chains of $20{,}000$ iterations ($5{,}000$ burn-in, thinning $5$) are run to examine convergence in detail. Convergence is assessed by the potential scale reduction factor $\widehat R$ \citep{gelman1992inference,vehtari2021rank} and the pooled effective sample size (ESS).

\paragraph{Performance measures metrics.} For each scalar parameter $\vartheta$ we report the bias and root mean squared error (RMSE) of the posterior mean over replicates, and the empirical coverage and mean width of the $95\%$ equal-tailed credible interval. For the time-indexed quantities $\mu_i$, $\log\sigma_i$ and $P_i(p)$, $p\in\{0.90,0.99\}$, the same measures are averaged over $i=1,\ldots,n$. Since $\Pr\{P_i(p)>P_{i'}(p)\mid\mathbf y\}=\Pr(\lambda_i<\lambda_{i'}\mid\mathbf y)$ for every $p$, we also assess the ordering of persistence across time points on $200$ random pairs $(i,i')$ per replicate, through the concordance of the posterior ordering with the true ordering and the calibration of the posterior ordering probabilities.

\subsection{Results}
\label{subsec:sim-results}

\subsubsection{Recovery of the shape and asymptotic persistence} 
Table~\ref{tab:sim-reps} reports the repeated-sampling performance for all model parameters, and Figure~\ref{fig:sim-reps}(a) shows the sampling distribution of the posterior means of $\xi$ and $P_\infty$. Under both models the common shape is recovered with small bias. Its RMSE decreases from 0.098 to 0.038 (GEV) and from 0.179 to 0.067 (GPD) as $n$ increases from $100$ to $500$, and the mean width of its $95\%$ credible interval shrinks from 0.392 to 0.168 and from 0.686 to 0.417. This is consistent with $\xi$ being pooled over all $n$ observations. The same holds for the asymptotic persistence: the credible intervals of $P_\infty$ have coverage 0.96 and 0.96 for the GEV model and 0.86 and 0.98 for the GPD model, at $n=100$ and $500$ respectively. The under-coverage of the GPD model at $n=100$ reflects a wide, right-skewed posterior for $\xi$ that is still partly shaped by the prior; it disappears at $n=500$. The GPD shape is less precisely estimated than the GEV shape at the same $n$. Each excess carries less information about the tail than a block maximum, and the latent scale effect is partially confounded with $\xi$ (see also Section~\ref{subsec:app-results}). For the regression coefficients, including those of the null covariate $x_2$, coverage lies between 0.88 and 1.00, close to the nominal level given the Monte Carlo standard error of about $0.03$ with $R=50$. The variance components $\tau_\mu^2$ and $\tau_\sigma^2$ are the exception, as anticipated: each latent effect $b_{ji}$ is informed by a single observation, so their credible intervals are wide and over-cover (e.g.\ mean width 4.759 for $\tau_\mu^2$ at $n=500$), and their posteriors largely reflect the prior. In each $(\text{model},n)$ setting, at least $93\%$ of the parameter--replicate combinations had $\widehat R<1.05$ with two chains; most exceptions involved the latent-location variance $\tau_\mu^2$ of the GEV model, and a few GPD replicates involved the pair $(\xi,\tau_\sigma^2)$, reflecting their partial confounding.

\begin{table}[t]
\centering
\small
\caption{Repeated-sampling performance over $R=50$ simulated data sets: bias and root mean squared error (RMSE) of the posterior mean, empirical coverage (Cov.) and mean width of the $95\%$ equal-tailed credible interval. $P_\infty=1/(1-\xi)$ for $\xi>0$ and $1$ otherwise; $P_Z(0.99)$ is the (time-invariant) GPD excess persistence~\eqref{eq:gpdPZ}.}
\label{tab:sim-reps}
\setlength{\tabcolsep}{4.5pt}
\begin{tabular}{@{}lr rrrr rrrr@{}}
\toprule
 & & \multicolumn{4}{c}{$n=100$} & \multicolumn{4}{c}{$n=500$} \\
\cmidrule(lr){3-6}\cmidrule(lr){7-10}
Parameter & True & Bias & RMSE & Cov. & Width & Bias & RMSE & Cov. & Width \\
\midrule
\multicolumn{10}{@{}l}{\textit{GEV model}} \\
$\beta_{\mu,0}$ & 40.000 & 0.412 & 1.308 & 0.92 & 4.869 & 0.161 & 0.655 & 0.90 & 2.081 \\
$\beta_{\mu,1}$ (time) & 2.000 & 0.189 & 1.097 & 0.94 & 4.419 & 0.176 & 0.614 & 0.88 & 1.926 \\
$\beta_{\mu,2}$ ($x_1$) & 4.000 & 0.163 & 1.179 & 0.92 & 4.575 & $-$0.084 & 0.498 & 0.94 & 1.865 \\
$\beta_{\mu,3}$ ($x_2$) & 0.000 & 0.089 & 1.150 & 0.90 & 4.389 & 0.064 & 0.592 & 0.88 & 1.903 \\
$\beta_{\sigma,0}$ & 2.303 & 0.009 & 0.099 & 0.98 & 0.463 & 0.003 & 0.048 & 0.96 & 0.192 \\
$\beta_{\sigma,1}$ (time) & 0.150 & $-$0.010 & 0.104 & 0.96 & 0.421 & 0.010 & 0.053 & 0.92 & 0.172 \\
$\beta_{\sigma,2}$ ($x_1$) & 0.100 & 0.020 & 0.097 & 0.94 & 0.429 & $-$0.005 & 0.043 & 0.96 & 0.171 \\
$\beta_{\sigma,3}$ ($x_2$) & 0.000 & $-$0.032 & 0.113 & 0.94 & 0.421 & 0.003 & 0.043 & 0.90 & 0.170 \\
$\tau_\mu^2$ & 1.000 & 0.408 & 1.073 & 1.00 & 8.582 & 0.055 & 0.258 & 1.00 & 4.759 \\
$\tau_\sigma^2$ & 0.040 & 0.005 & 0.018 & 1.00 & 0.140 & $-$0.002 & 0.014 & 1.00 & 0.078 \\
$\xi$ & 0.200 & 0.031 & 0.098 & 0.96 & 0.392 & 0.000 & 0.038 & 0.96 & 0.168 \\
$P_\infty$ & 1.250 & 0.106 & 0.226 & 0.96 & 0.890 & 0.007 & 0.060 & 0.96 & 0.271 \\
\addlinespace
\multicolumn{10}{@{}l}{\textit{GPD model}} \\
$\beta_{\sigma,0}$ & 0.500 & $-$0.023 & 0.199 & 0.92 & 0.692 & $-$0.000 & 0.078 & 1.00 & 0.339 \\
$\beta_{\sigma,1}$ (time) & 0.200 & 0.005 & 0.126 & 0.94 & 0.505 & 0.005 & 0.047 & 1.00 & 0.224 \\
$\beta_{\sigma,2}$ ($x_1$) & 0.150 & 0.026 & 0.132 & 0.94 & 0.516 & $-$0.009 & 0.052 & 0.96 & 0.224 \\
$\beta_{\sigma,3}$ ($x_2$) & 0.000 & $-$0.004 & 0.133 & 0.90 & 0.506 & 0.010 & 0.055 & 0.94 & 0.225 \\
$\tau_\sigma^2$ & 0.090 & 0.015 & 0.016 & 1.00 & 0.325 & 0.017 & 0.025 & 1.00 & 0.353 \\
$\xi$ & 0.250 & $-$0.013 & 0.179 & 0.86 & 0.686 & $-$0.016 & 0.067 & 0.98 & 0.417 \\
$P_\infty$ & 1.333 & 0.179 & 0.448 & 0.86 & 1.545 & 0.007 & 0.109 & 0.98 & 0.605 \\
$P_Z(0.99)$ & 1.487 & 0.171 & 0.431 & 0.86 & 1.524 & 0.008 & 0.098 & 0.98 & 0.566 \\
\bottomrule
\end{tabular}
\end{table}

\subsubsection{Recovery of finite-level persistence} Table~\ref{tab:sim-latent} summarizes recovery of the time-indexed quantities. At $n=500$, the pointwise $95\%$ credible intervals of $\mu_i$, $\log\sigma_i$ and $P_i(p)$ have coverage between $0.93$ and $0.98$; at $n=100$, coverage is slightly below nominal for the GPD level persistence ($0.88$ for $P_i(0.99)$), in line with the behaviour of $\xi$. The RMSE of $\widehat P_i(0.99)$ falls from 0.135 to 0.042 (GEV) and from 0.235 to 0.073 (GPD). The correlation between $\widehat P_i(p)$ and its true value is moderate (0.42 and 0.64 at $n=500$). Part of the variation of $P_i(p)$ across $i$ is due to the observation-level latent effects, which a single observation cannot resolve. The systematic, covariate-driven part is recovered well: Figure~\ref{fig:sim-rep1}(a) shows, for the representative data set, that the posterior median of $P_i(0.99)$ tracks the upward trend of the true trajectory, and that the $95\%$ bands contain the true values. In this data set, the posterior probability that the mean persistence over the last $50$ time points exceeds that over the first $50$ is 1.000 (GEV) and 1.000 (GPD), in agreement with the data-generating trend.

\begin{table}[t]
\centering
\small
\caption{Recovery of the time-indexed quantities, averaged over the $n$ time points and $R=50$ replicates: RMSE of the posterior mean, coverage and mean width of pointwise $95\%$ credible intervals, and correlation (Cor.) between posterior mean and truth. For the GPD, $P_i(p)$ is the level persistence $P_{Y,i}(p)$ of~\eqref{eq:gpdPYdraw} with $v=10$. Ordering concordance is the proportion of $200$ random pairs $(i,i')$ for which $\Pr\{P_i(p)>P_{i'}(p)\mid\mathbf y\}>0.5$ agrees with the true ordering.}
\label{tab:sim-latent}
\setlength{\tabcolsep}{5pt}
\begin{tabular}{@{}l rrrr rrrr@{}}
\toprule
 & \multicolumn{4}{c}{$n=100$} & \multicolumn{4}{c}{$n=500$} \\
\cmidrule(lr){2-5}\cmidrule(lr){6-9}
Quantity & RMSE & Cov. & Width & Cor. & RMSE & Cov. & Width & Cor. \\
\midrule
\multicolumn{9}{@{}l}{\textit{GEV model}} \\
$\mu_i$ & 2.438 & 0.95 & 9.836 & 0.92 & 1.513 & 0.93 & 5.517 & 0.96 \\
$\log\sigma_i$ & 0.276 & 0.96 & 1.152 & 0.50 & 0.216 & 0.93 & 0.816 & 0.65 \\
$P_i(0.90)$ & 0.119 & 0.94 & 0.585 & 0.26 & 0.048 & 0.95 & 0.206 & 0.43 \\
$P_i(0.99)$ & 0.135 & 0.96 & 0.722 & 0.26 & 0.042 & 0.96 & 0.208 & 0.42 \\
\multicolumn{9}{@{}l}{\quad Ordering concordance of $\Pr\{P_i(p)>P_{i'}(p)\mid\mathbf y\}$: 0.58 ($n=100$), 0.64 ($n=500$)} \\
\addlinespace
\multicolumn{9}{@{}l}{\textit{GPD model}} \\
$\log\sigma_i$ & 0.413 & 0.94 & 1.624 & 0.52 & 0.314 & 0.96 & 1.353 & 0.64 \\
$P_i(0.90)$ & 0.155 & 0.91 & 0.796 & 0.52 & 0.067 & 0.97 & 0.330 & 0.64 \\
$P_i(0.99)$ & 0.235 & 0.88 & 1.230 & 0.52 & 0.073 & 0.98 & 0.432 & 0.64 \\
\multicolumn{9}{@{}l}{\quad Ordering concordance of $\Pr\{P_i(p)>P_{i'}(p)\mid\mathbf y\}$: 0.69 ($n=100$), 0.71 ($n=500$)} \\
\bottomrule
\end{tabular}
\end{table}

\subsubsection{Ordering of persistence} 
\begin{figure}[t]
\centering
\includegraphics[width=\textwidth]{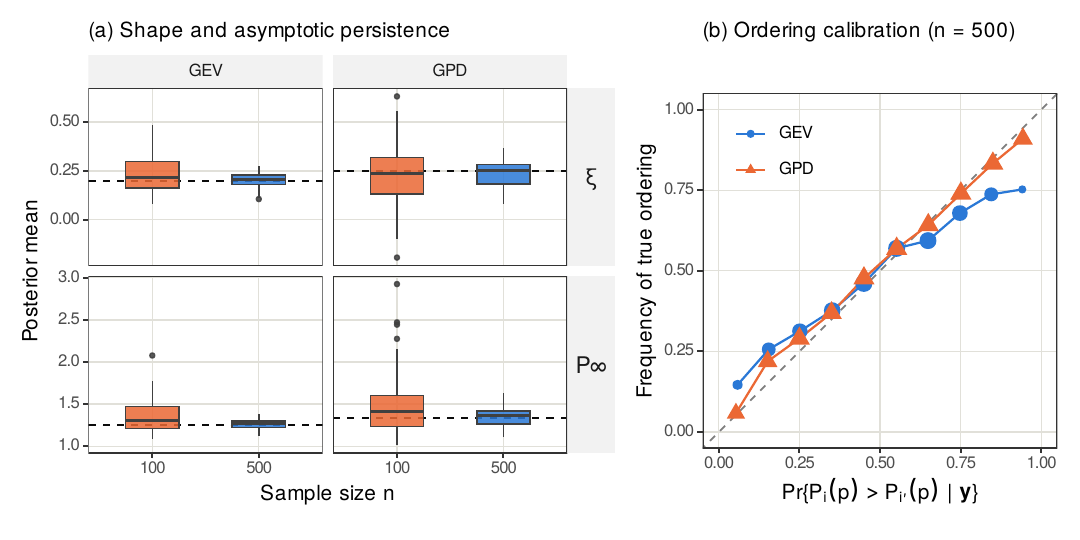}
\caption{Repeated-sampling results ($R=50$). (a)~Posterior means of the common shape $\xi$ and of the asymptotic persistence $P_\infty$ for $n=100$ and $500$; dashed lines mark the true values. (b)~Calibration of the posterior ordering probabilities $\Pr\{P_i(p)>P_{i'}(p)\mid\mathbf y\}$ at $n=500$: the frequency with which the ordering holds for the true parameters, against the posterior probability binned in deciles; point size is proportional to the square root of the bin count.}
\label{fig:sim-reps}
\end{figure}

Because $\Pr\{P_i(p)>P_{i'}(p)\mid\mathbf y\}=\Pr(\lambda_i<\lambda_{i'}\mid\mathbf y)$ for every $p$, a single posterior probability orders two time points at all levels. The ordering implied by these probabilities agrees with the true ordering for a proportion 0.64 (GEV) and 0.71 (GPD) of the random pairs at $n=500$. Many pairs differ only through their latent effects and are close to indistinguishable. Figure~\ref{fig:sim-reps}(b) shows that, under the GPD model, the ordering probabilities are well calibrated: an ordering assigned posterior probability $q$ holds in a proportion close to $q$ of the pairs, with a maximum absolute deviation of 0.07 over the deciles. Under the GEV model, calibration is good for intermediate probabilities but the extreme deciles are overconfident (maximum deviation 0.19). We attribute this to the weak identification of the location latent effects noted above; it supports using covariate-only location channels when each time point contributes a single maximum, as we do in Section~\ref{sec:application}.

\subsubsection{A representative data set}  
\begin{figure}[t]
\centering
\includegraphics[width=\textwidth]{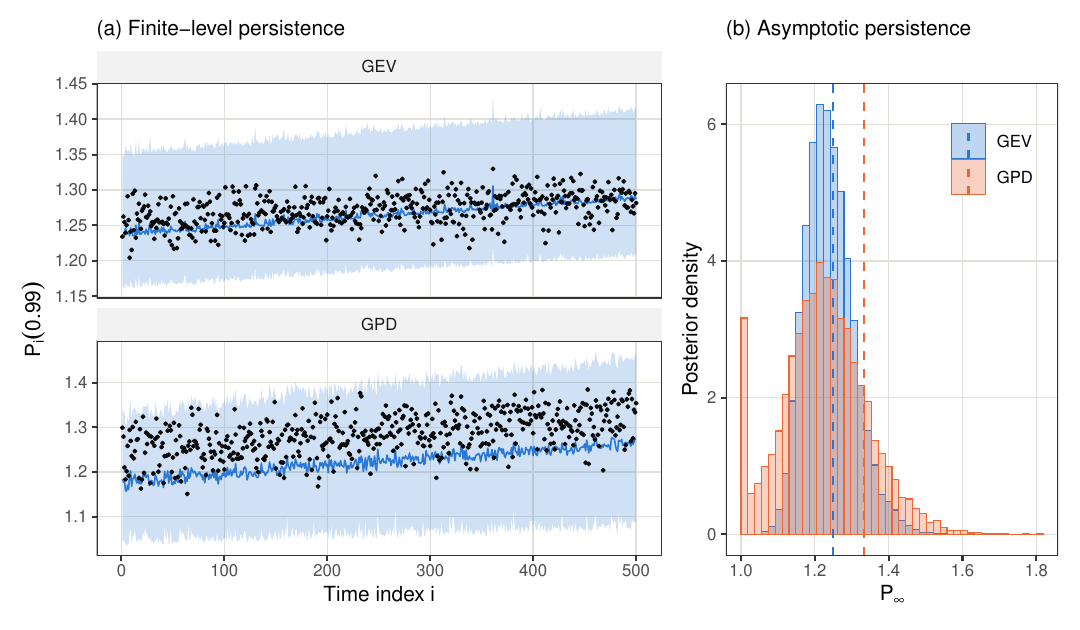}
\caption{Representative simulated data set ($n=500$). (a)~True finite-level persistence $P_i(0.99)$ (points), posterior median (line) and $95\%$ pointwise credible band; for the GPD, $P_i(0.99)$ is the level persistence $P_{Y,i}(0.99)$. True values outside the band are shown in orange. (b)~Posterior distributions of $P_\infty$; dashed lines mark the true values $1.25$ (GEV) and $4/3$ (GPD).}
\label{fig:sim-rep1}
\end{figure}

For the data set with $n=500$ analysed with four chains, all parameters have $\widehat R\le1.018$ and pooled ESS of at least 1,407 (Table~\xref{S-tab:sim-rep1} and Figure~\xref{S-fig:sim-traces} in the Supplementary Material). The posterior mean and $95\%$ credible interval of $\xi$ are $0.191\ [0.115, 0.280]$ for the GEV model and $0.171\ [-0.092, 0.319]$ for the GPD model, and those of $P_\infty$ are $1.240\ [1.130, 1.389]$ and $1.226\ [1.000, 1.468]$. For the GPD model, $\Pr(\xi>0\mid\mathbf y)=0.950$. The remaining posterior mass at $\xi\le0$ appears in Figure~\ref{fig:sim-rep1}(b) as an atom of $P_\infty$ at one, in line with Remark~\ref{rem:regularVariationCaveat}.

\section{Application to England and Wales precipitation extremes}
\label{sec:application}

\subsection{Data}
\label{subsec:app-data}

We analyse monthly precipitation totals (mm) for England and Wales from the homogenized England and Wales Precipitation (EWP) series of the UK Met Office Hadley Centre \citep{MetOfficeHadUKP}, over January 1950 to December 2025 ($912$ months). Three indices of large-scale climate variability are used as covariates: the El Ni\~no--Southern Oscillation, through the Ni\~no~3.4 sea surface temperature anomaly; the Indian Ocean Dipole, through the Dipole Mode Index \citep[DMI;][]{saji1999dipole}, both computed from HadISST1.1 \citep{rayner2003hadisst} and distributed by the NOAA Physical Sciences Laboratory \citep{NOAAPSL}; and the North Atlantic Oscillation index of the NOAA Climate Prediction Center \citep{NOAACPC}. The start year 1950 is the first year for which all four series are available.

\begin{figure}[t]
\centering
\includegraphics[width=\textwidth]{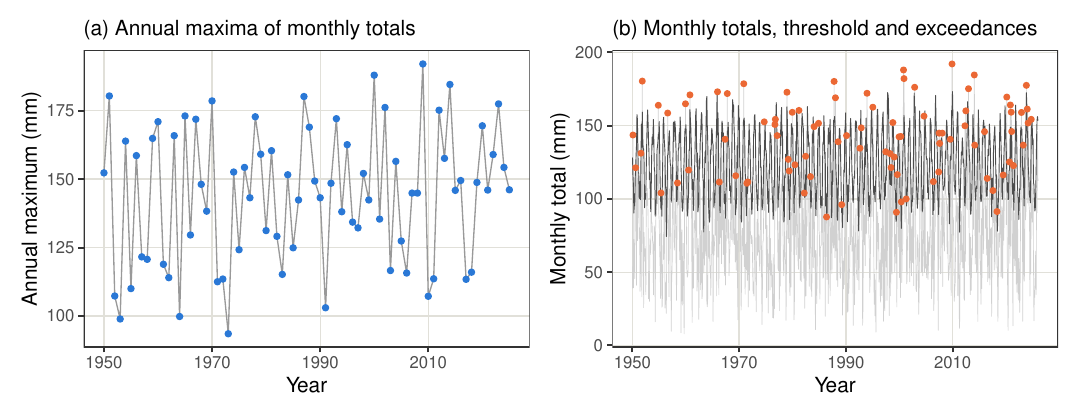}
\caption{England and Wales precipitation, 1950--2025. (a)~Annual maxima of monthly totals. (b)~Monthly totals (grey), the $0.90$ quantile-regression threshold (dark grey) and the $88$ threshold exceedances (orange).}
\label{fig:app-data}
\end{figure}

Two data sets are constructed from these records (Figure~\ref{fig:app-data}).
\begin{itemize}
\item \emph{Block maxima.} For each year $i=1950,\ldots,2025$ we take the largest monthly total, $Y_i$, giving $n=76$ annual maxima, together with the annual means of the three indices. Most annual maxima occur between October and January.
\item \emph{Threshold excesses.} A time-varying threshold $v_t$ is estimated by linear quantile regression \citep{koenker1978regression,northrop2011threshold} of the monthly totals at level $0.90$, with a first-harmonic seasonal cycle $(\sin(2\pi m_t/12),\cos(2\pi m_t/12))$, where $m_t$ is the calendar month, and the three monthly indices as regressors. Months with $y_t>v_t$ give $n=88$ excesses $z_t=y_t-v_t$. The six observations that the quantile-regression fit interpolates exactly ($z_t=0$) are excluded.
\end{itemize}

\subsection{Model specification}
\label{subsec:app-model}

For both models the covariate vector is $\mathbf x_i=(1,\,\mathrm{year}_i,\,\mathrm{ENSO}_i,\,\mathrm{IOD}_i,\,\mathrm{NAO}_i)^\top$, with every covariate standardized. The linear time term allows a gradual trend in the location and scale, and hence, by Proposition~\ref{prop:locScale}, in the finite-level persistence. For the GEV model the covariates enter both the location and the log-scale; for the GPD model they enter the log-scale of the excesses. With a single maximum per year and $n<100$, observation-specific latent effects $b_{ji}$ are only weakly identified (Section~\ref{subsec:sim-results}). In particular, a location latent effect in the GEV model allows the scale to collapse while $\tau_\mu^2$ absorbs the variability of the maxima. We therefore use covariate-only channels ($b_{ji}\equiv0$) and report the latent-effect fits as a sensitivity analysis (Section~\xref{S-sec:supp_app}).

\begin{figure}[ht]
\centering
\includegraphics[width=0.8\textwidth]{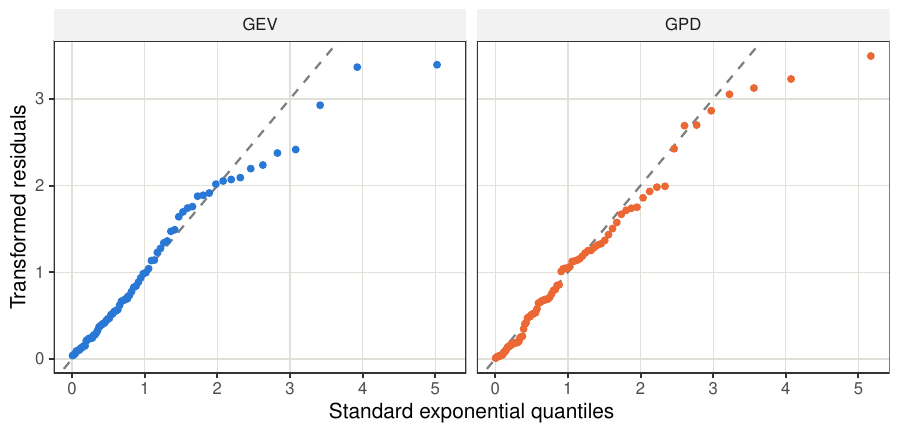}
\caption{Exponential quantile--quantile plots of the transformed residuals $-\log\{1-\widehat F_i(y_i)\}$ for the GEV model of annual maxima and the GPD model of threshold excesses (England and Wales precipitation, 1950--2025), evaluated at posterior mean parameters.}
\label{fig:app-qq}
\end{figure}

The shape parameter is common to all observations and linked through~\eqref{eq:boundedXiLink}, with bounds $\xi_{\min}=-1$ and $\xi_{\max}=0.8$ and prior $\widetilde\xi\sim N(0,1.7^2)$. The interval admits the Weibull, Gumbel and Fr\'echet domains, so the data determine the sign of $\xi$. The regression priors are $\boldsymbol\beta_\mu\sim N(\mathbf 0,10^6 I)$ and $\boldsymbol\beta_\sigma\sim N(\mathbf 0,10I)$ for the GEV model, and $\boldsymbol\beta_\sigma\sim N(\mathbf 0,100I)$ for the GPD model. For each model, four chains are run from dispersed starting values for $40{,}000$ iterations; the first $10{,}000$ are discarded as burn-in and every tenth draw is retained, giving $12{,}000$ posterior draws. All parameters have $\widehat R\le1.001$ and pooled ESS above $8{,}900$. Model fit is assessed through the transformed residuals $-\log\{1-\widehat F_i(y_i)\}$, evaluated at the posterior mean parameters, which are approximately standard exponential under a correctly specified model. The exponential quantile--quantile plots in Figure~\ref{fig:app-qq} show no lack of fit for either model, in agreement with Kolmogorov--Smirnov $p$-values of $0.99$ for the GEV and $0.32$ for the GPD model.

For the GPD model, finite-level persistence is reported both for the excesses, $P_Z(p)$ in~\eqref{eq:gpdPZdraw}, and for the rainfall level given an exceedance, $P_{Y,i}(p)$ in~\eqref{eq:gpdPYdraw}, with $\lambda_i=v_i/\sigma_i$ built from the quantile-regression threshold. Because $v_i$ moves with season and climate state, only $P_{Y,i}(p)$ carries covariate-driven variation in the relative escalation of rainfall.

\subsection{Results}
\label{subsec:app-results}

\begin{table}[t]
\centering
\small
\caption{Posterior summaries for the England and Wales precipitation extremes (1950--2025): posterior means and $95\%$ credible intervals for the GEV model of annual maxima ($n=76$) and the GPD model of threshold excesses ($n=88$). Covariates are standardized. Across all parameters, $\widehat R\le 1.001$ and the pooled ESS is at least 8,910.}
\label{tab:app-post}
\setlength{\tabcolsep}{4pt}
\begin{tabular}{@{}l rc@{\hspace{1.2em}} rc@{}}
\toprule
 & \multicolumn{2}{c}{GEV (annual maxima)} & \multicolumn{2}{c}{GPD (excesses)} \\
\cmidrule(lr){2-3}\cmidrule(lr){4-5}
 & Mean & $95\%$ CI & Mean & $95\%$ CI \\
\midrule
\multicolumn{5}{@{}l}{\textit{Location, $\boldsymbol\beta_\mu$}} \\
\quad Intercept & 135.804 & [128.896, 142.659] & \multicolumn{2}{c}{--} \\
\quad Year (trend) & 6.217 & [$-$0.850, 13.101] & \multicolumn{2}{c}{--} \\
\quad ENSO & 2.579 & [$-$3.714, 8.774] & \multicolumn{2}{c}{--} \\
\quad IOD & $-$4.205 & [$-$11.185, 2.939] & \multicolumn{2}{c}{--} \\
\quad NAO & 0.766 & [$-$5.848, 7.226] & \multicolumn{2}{c}{--} \\
\multicolumn{5}{@{}l}{\textit{Log-scale, $\boldsymbol\beta_\sigma$}} \\
\quad Intercept & 3.225 & [3.035, 3.439] & 3.291 & [2.976, 3.579] \\
\quad Year (trend) & $-$0.136 & [$-$0.349, 0.059] & 0.127 & [$-$0.030, 0.260] \\
\quad ENSO & $-$0.041 & [$-$0.226, 0.139] & $-$0.037 & [$-$0.204, 0.147] \\
\quad IOD & 0.154 & [$-$0.074, 0.380] & $-$0.015 & [$-$0.202, 0.178] \\
\quad NAO & $-$0.053 & [$-$0.253, 0.136] & 0.052 & [$-$0.081, 0.182] \\
\multicolumn{5}{@{}l}{\textit{Tail index and persistence}} \\
\quad $\xi$ & $-$0.310 & [$-$0.517, $-$0.086] & $-$0.471 & [$-$0.728, $-$0.184] \\
\quad $P_\infty$ & 1.000 & [1.000, 1.000] & 1.000 & [1.000, 1.000] \\
\quad $P_Z(0.99)$ & \multicolumn{2}{c}{--} & 1.047 & [1.015, 1.117] \\
\quad $\Pr(\xi>0\mid\mathbf y)$ & \multicolumn{2}{c}{0.004} & \multicolumn{2}{c}{0.003} \\
\addlinespace
\multicolumn{5}{@{}l}{\textit{Posterior probability of a positive effect, $\Pr(\beta>0\mid\mathbf y)$ (Year, ENSO, IOD, NAO)}} \\
\quad Location (GEV) & \multicolumn{2}{c}{0.96, 0.80, 0.12, 0.60} & \multicolumn{2}{c}{--} \\
\quad Log-scale & \multicolumn{2}{c}{0.09, 0.32, 0.91, 0.30} & \multicolumn{2}{c}{0.95, 0.33, 0.43, 0.79} \\
\bottomrule
\end{tabular}
\end{table}

Table~\ref{tab:app-post} summarizes the posterior distributions. The location of the annual maxima increases over time, with posterior mean $6.2$\,mm per standard deviation of the year (about $22$ years) and $\Pr(\beta_{\mu,\mathrm{year}}>0\mid\mathbf y)=0.96$. The scale of the annual maxima decreases slightly ($\Pr(\beta_{\sigma,\mathrm{year}}>0\mid\mathbf y)=0.09$). In contrast, the scale of the threshold excesses increases ($\Pr(\beta_{\sigma,\mathrm{year}}>0\mid\mathbf y)=0.95$). None of the climate indices has a credible effect on either model once the trend is included. The largest posterior probabilities are $0.91$ for a positive IOD effect on the GEV scale and $0.80$ for a positive ENSO effect on the GEV location.

\subsubsection{Asymptotic persistence}  Both models place almost all posterior mass on a negative shape parameter: $\xi$ has posterior mean $-0.310$ (95\% CI $[-0.517,-0.086]$) under the GEV model and $-0.471$ ($[-0.728,-0.184]$) under the GPD model. The posterior probabilities $\Pr(\xi>0\mid\mathbf y)$ are $0.004$ and $0.003$. Monthly precipitation totals in England and Wales therefore lie in the Weibull domain of attraction, with a finite upper endpoint. In line with Remark~\ref{rem:regularVariationCaveat}, the asymptotic persistence equals one with posterior probability above $0.99$: far enough into the tail, exceedances of a quantile are not, on average, proportionally larger than the quantile itself. This is consistent with the aggregation of daily rainfall into monthly totals, which shortens the tail. The fits with latent effects give the same conclusion, with $\xi$ equal to $-0.331$ $[-0.590,-0.073]$ (GEV, latent scale effect) and $-0.611$ $[-0.945,-0.238]$ (GPD, latent scale effect). The more negative value in the GPD case reflects the partial confounding between $\xi$ and $\tau_\sigma^2$: a log-normal mixture of GPD scales has a heavier tail than its components.

\subsubsection{Finite-level persistence}  
\begin{figure}[t]
\centering
\includegraphics[width=\textwidth]{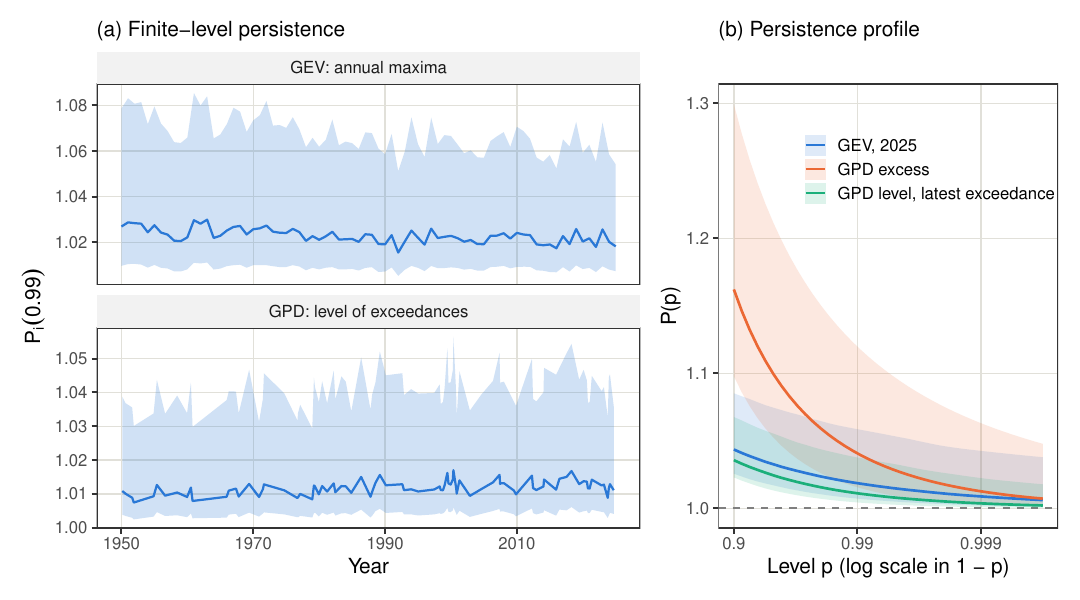}
\caption{Persistence of England and Wales precipitation extremes. (a)~Posterior median and $95\%$ pointwise credible band of $P_i(0.99)$ for the annual maxima (GEV) and of the level persistence $P_{Y,i}(0.99)$ of the threshold exceedances (GPD). (b)~Persistence profiles $p\mapsto P(p)$ with $95\%$ credible bands for the GEV annual maximum in 2025, the GPD excess, and the GPD level of the latest exceedance; the dashed line is the common asymptotic value $P_\infty=1$.}
\label{fig:app-persistence}
\end{figure}

Persistence at finite levels is informative even though $P_\infty=1$ (Figure~\ref{fig:app-persistence}). The excess persistence is $P_Z(0.99)=1.047$ $[1.015,1.117]$. Hence the mean excess of a monthly total beyond the $99$th percentile of the excess distribution is about $5\%$ of that percentile. The persistence profiles in Figure~\ref{fig:app-persistence}(b) decrease towards one as $p\uparrow1$, as Propositions~\ref{prop:gpdPersistence}(d) and~\ref{prop:gevPersistence}(d) predict for $\xi<0$. For the annual maxima, the posterior median of $P_i(0.99)$ ranges from $1.016$ to $1.030$ over 1950--2025 ($1.036$ to $1.074$ at $p=0.9$). For the rainfall level given an exceedance, $P_{Y,i}(0.99)$ ranges from $1.007$ to $1.017$.

The two series move in opposite directions over time. For the annual maxima, the location grows while the scale does not, so $\lambda_i=\mu_i/\sigma_i$ increases and persistence decreases. The posterior probability that mean persistence over 1996--2025 exceeds that over 1950--1979 is $0.12$, with difference $-0.004$ $[-0.016,0.002]$. For the exceedances, the typical excess size $\sigma_i$ grows relative to the threshold, so $\lambda_i=v_i/\sigma_i$ decreases and persistence increases. The corresponding probability is $0.97$, with difference $0.003$ $[-0.000,0.008]$. Thus wetter annual maxima in recent decades reflect an upward shift rather than a wider spread. In contrast, months that exceed the seasonal threshold now overshoot it by relatively more. By~\eqref{eq:persistenceContrast}, both statements hold at every level $p$. The changes are small in absolute terms, which is expected for a bounded upper tail.

\section{Conclusion and future research directions}
\label{sec:conclusion}

This paper treats tail persistence as the primary inferential target in extreme value analysis. Under regular variation, the asymptotic persistence is determined by the extreme value index alone, $P_\infty=1/(1-\xi)$. At finite levels, persistence has closed forms under the GPD and GEV models and, for a location--scale family with a common shape, depends on the location and scale only through their ratio $\lambda=\mu/\sigma$. These results motivate a Bayesian hierarchical model in which the shape is common across time, while the location and scale vary with covariates and latent effects. The model separates a single, well-identified asymptotic persistence from a time-varying finite-level persistence. It also reduces comparisons of persistence across time to comparisons of $\lambda$, which hold simultaneously at every level $p$.

The simulation study shows that the common shape and the asymptotic persistence are recovered with small bias and close to nominal coverage under both the GEV and GPD formulations, and that their precision improves markedly with sample size. The finite-level persistence trajectories are estimated with near-nominal pointwise coverage, and the posterior probabilities of an ordering in persistence between time points are well calibrated. Variance components of observation-level latent effects are weakly identified when each time point contributes one observation, and their posteriors should be read together with the prior. In the application to England and Wales precipitation, both models place the upper tail of monthly totals in the Weibull domain, so that $P_\infty=1$. The finite-level persistence remains informative, and it changes in opposite directions for annual maxima and for threshold exceedances. This illustrates how persistence complements return levels and tail indices.

Several extensions are natural. First, the regime-wise shape parameters of Remark~\ref{rem:regimeShape} allow formal comparison of asymptotic persistence before and after a known change point. Second, spatial or spatio-temporal latent effects, for example Gaussian Markov random fields in the location and scale channels, would yield maps of finite-level persistence and of the probability of its increase. Third, a joint model for the exceedance probability and the excess distribution would give persistence at unconditional levels of the parent distribution, as noted after~\eqref{eq:gpdPYdraw}. Finally, the identifiability of latent effects at the observation level, and the confounding between $\xi$ and the latent scale variance in the GPD model, deserve further study, for example through penalized-complexity priors on $\tau_j^2$ \citep{Simpson2017}. The open-source \texttt{R} package accompanying this paper implements both models and reproduces the simulation study and the application, so that the methodology can be applied directly to other data sets.

\newpage
\section*{Acknowledgement}
The authors acknowledge the use of ChatGPT and Claude to assist with language refinement, phrasing, and improving the clarity and readability of certain parts of the manuscript. The original research idea, methodological development, theoretical contributions, analyses, and scientific interpretations are the authors' own. The authors take full responsibility for the content and conclusions of the manuscript. 
\section*{Data and code availability}
The England and Wales Precipitation series is available from the UK Met Office Hadley Centre \citep{MetOfficeHadUKP}, the HadISST-based Ni\~no~3.4 and Dipole Mode indices from the NOAA Physical Sciences Laboratory \citep{NOAAPSL}, and the NAO index from the NOAA Climate Prediction Center \citep{NOAACPC}. The analysis data sets were constructed by merging the monthly series by year and month for January 1950 to December 2025. The processed data sets and the R code for the sampler, the simulation study and the application are available from the corresponding author upon reasonable request.

We have implemented our methodology in an open-source \texttt{R} package hosted on GitHub, which will be released for public at the final stage of the publication process. 

\section*{Supplementary Information}
Supplementary material for this article is available at the following link \href{https://drive.google.com/file/d/1y7sSVbI9A2kZzDKPTIupEbKFMF3v8hi2/view?usp=sharing}{SupplementaryInformation}

\section*{Author contributions}
Nikita (mostly leading the project),  and Shiv developed the theoretical results and Bayesian methodology, implemented the computational algorithms, performed the simulation studies and data analysis, and drafted the manuscript. Dreamlee Sharma contributed to the discussion and refinement of the original idea, literature review, review of the manuscript, and project supervision. Rishikesh Yadav provided overall supervision throughout the study, including the formulation of the original idea, model development, computational methodology, and review and editing of the manuscript. All authors read and approved the final manuscript.

\bibliographystyle{plainnat}
\bibliography{ref}

\appendix

\section{Proofs of the main results}
\label{app:proofs}

This appendix contains the proofs of the results of Section~\ref{sec:methodology}, together with two further results used there: the ordering identity~\eqref{eq:persistenceContrast} (Proposition~\ref{prop:contrast}) and persistence at unconditional levels of the parent distribution (Proposition~\ref{prop:uncond}). Background on regular variation is given in Section~\xref{S-sec:supp_rv} of the Supplementary Material.

\subsection{Regular variation of the upper quantile function}
\label{app:quantileRV}

\begin{lemma}
\label{lem:quantileRV}
If $\bar F\in RV_{-1/\xi}$ with $\xi>0$, then there is a slowly varying function $L_1$ such that
\begin{equation}
Q(1-t)=t^{-\xi}L_1(1/t),\qquad t\downarrow0,
\label{eq:quantileRV}
\end{equation}
that is, $t\mapsto Q(1-t)$ is regularly varying at zero with index $-\xi$.
\end{lemma}

\begin{proof}
Write $\bar F(x)=x^{-1/\xi}L(x)$ with $L$ slowly varying, so that $1/\bar F\in RV_{1/\xi}$. The tail quantile function $U(s):=Q(1-1/s)$, $s>1$, is the left-continuous generalized inverse of $1/\bar F$. By the inversion theorem for regularly varying functions \citep[Theorem~1.5.12]{bingham1989regular} (see also \citealp[Appendix~B.1]{deHaan2006}), the generalized inverse of a function in $RV_{1/\xi}$, $1/\xi>0$, belongs to $RV_{\xi}$. Hence $U(s)=s^{\xi}L_1(s)$ for some slowly varying $L_1$, and setting $s=1/t$ gives~\eqref{eq:quantileRV}.
\end{proof}

\subsection{Proof of Theorem~\ref{thm:persistence}}
\label{app:proofThm}

\begin{proof}
Let $u=1-t$ with $t\downarrow0$. From the definition~\eqref{eq:persistence} and the change of variables $y=1-p$,
\[
P(1-t)
=\frac{1}{t\,Q(1-t)}\int_{1-t}^{1}Q(p)\,dp
=\frac{1}{t\,Q(1-t)}\int_{0}^{t}Q(1-y)\,dy.
\]
By Lemma~\ref{lem:quantileRV}, $Q(1-y)=y^{-\xi}L_1(1/y)$, so that
\begin{equation}
P(1-t)=\frac{\int_0^t y^{-\xi}L_1(1/y)\,dy}{t^{1-\xi}L_1(1/t)}.
\label{eq:beforeKaramata}
\end{equation}
The integrand is regularly varying at zero with index $-\xi>-1$. Karamata's theorem \citep[Theorem~1.5.11]{bingham1989regular}, applied to $x\mapsto x^{\xi-2}L_1(x)\in RV_{\xi-2}$ after the substitution $y=1/x$, therefore gives
\[
\int_0^t y^{-\xi}L_1(1/y)\,dy\;\sim\;\frac{t^{1-\xi}L_1(1/t)}{1-\xi},\qquad t\downarrow0,
\]
and in particular the integral is finite for small $t$. Substituting into~\eqref{eq:beforeKaramata} yields $P(1-t)\to1/(1-\xi)$ as $t\downarrow0$, that is, $\lim_{u\uparrow1}P(u)=1/(1-\xi)$.
\end{proof}

\subsection{Verification of regular variation for the GPD and GEV}
\label{app:verifyRV}

For the standard GPD with $\xi>0$, $Q_W(1-t)=(t^{-\xi}-1)/\xi=t^{-\xi}L_1(1/t)$ with $L_1(x)=(1-x^{-\xi})/\xi\to1/\xi$, which is slowly varying. For the standard GEV with $\xi>0$, $\ell_{1-t}=-\log(1-t)=t\{1+t/2+O(t^2)\}$, so $Q_W(1-t)=(\ell_{1-t}^{-\xi}-1)/\xi=t^{-\xi}L_1(1/t)$ with $L_1(x)\to1/\xi$, again slowly varying. In both cases Lemma~\ref{lem:quantileRV} and Theorem~\ref{thm:persistence} apply, in agreement with the limits in Propositions~\ref{prop:gpdPersistence}(d) and~\ref{prop:gevPersistence}(d).

\subsection{Proof of Lemma~\ref{lem:meanExcess}}
\label{app:proofLemma}

\begin{proof}
Let $U\sim\mathrm{Unif}(0,1)$. Then $Y\overset{d}{=}Q(U)$ and, since $Q$ is strictly increasing, $\{Q(U)>Q(p)\}=\{U>p\}$, an event of probability $1-p$. Hence
\[
E[Y\mid Y>Q(p)]
=\frac{E\left[Q(U)\,\mathbf 1\{U>p\}\right]}{1-p}
=\frac{1}{1-p}\int_p^1Q(s)\,ds
=m(p),
\]
which is finite because $E|Y|<\infty$. Subtracting $Q(p)$ gives $e(p)=E[Y-Q(p)\mid Y>Q(p)]\ge0$. Finally, by~\eqref{eq:persistence}, $P(p)=m(p)/Q(p)=\{Q(p)+e(p)\}/Q(p)$.
\end{proof}

\subsection{Proof of Proposition~\ref{prop:locScale}}
\label{app:proofLocScale}

\begin{proof}
Since $\sigma>0$, $Q_Y(p)=\mu+\sigma Q_W(p)$, and therefore
\[
m_Y(p)=\frac{1}{1-p}\int_p^1\{\mu+\sigma Q_W(s)\}\,ds=\mu+\sigma\, m_W(p),
\qquad
e_Y(p)=m_Y(p)-Q_Y(p)=\sigma\, e_W(p).
\]
By Lemma~\ref{lem:meanExcess},
\[
P_Y(p)
=1+\frac{e_Y(p)}{Q_Y(p)}
=1+\frac{\sigma e_W(p)}{\mu+\sigma Q_W(p)}
=1+\frac{e_W(p)}{\mu/\sigma+Q_W(p)},
\]
which is~\eqref{eq:PlocScale}, and (i) follows. Differentiating~\eqref{eq:PlocScale} with respect to $\lambda$ on $\{\lambda>-Q_W(p)\}$ gives~\eqref{eq:dPdlambda}, hence (ii). Setting $\lambda=0$ gives $P_Y(p)=1+e_W(p)/Q_W(p)=P_W(p)$ by Lemma~\ref{lem:meanExcess}, hence (iii). For (iv), write
\[
P_Y(p)-1
=\frac{e_W(p)}{Q_W(p)}\cdot\frac{Q_W(p)}{\lambda+Q_W(p)}
=\{P_W(p)-1\}\cdot\frac{Q_W(p)}{\lambda+Q_W(p)}.
\]
As $p\uparrow1$, the second factor tends to one because $Q_W(p)\to\infty$, while $P_W(p)\to1/(1-\xi)$ by Theorem~\ref{thm:persistence}, since $\bar F_W\in RV_{-1/\xi}$ for the Fr\'echet-domain families considered (Section~\ref{app:verifyRV}). Hence $P_Y(p)\to1/(1-\xi)$.
\end{proof}

\subsection{Proof of Proposition~\ref{prop:gpdPersistence}}
\label{app:proofGPD}

\begin{proof}
(a) For $\xi\ne0$, inverting~\eqref{eq:gpdCDF} with $\sigma=1$ gives $Q_W(p)=\{(1-p)^{-\xi}-1\}/\xi=(a_p-1)/\xi$. Since $\xi<1$,
\[
\int_p^1(1-s)^{-\xi}\,ds=\frac{(1-p)^{1-\xi}}{1-\xi},
\]
so that
\[
m_W(p)
=\frac{1}{1-p}\int_p^1\frac{(1-s)^{-\xi}-1}{\xi}\,ds
=\frac{1}{\xi}\left\{\frac{a_p}{1-\xi}-1\right\},
\]
\[
e_W(p)=m_W(p)-Q_W(p)
=\frac{1}{\xi}\left\{\frac{a_p}{1-\xi}-a_p\right\}
=\frac{a_p}{1-\xi},
\]
which is~\eqref{eq:gpdMeanExcess}. Equivalently, $e_W(p)$ is the GPD mean residual life $(1+\xi z)/(1-\xi)$ evaluated at $z=Q_W(p)$. For $\xi=0$, $W$ is standard exponential, $Q_W(p)=-\log(1-p)$, and by memorylessness $e_W(p)=1$, which is the value of $a_p/(1-\xi)$ at $\xi=0$.

(b) $Z=\sigma W$ is a pure scale transformation, so by Proposition~\ref{prop:locScale}(iii), $P_Z(p)=P_W(p)=1+e_W(p)/Q_W(p)$. For $\xi\ne0$,
\[
P_Z(p)
=1+\frac{\xi a_p}{(1-\xi)(a_p-1)}
=1+\frac{\xi}{1-\xi}\left\{1+\frac{1}{a_p-1}\right\}
=\frac{1}{1-\xi}+\frac{\xi}{(1-\xi)(a_p-1)},
\]
and for $\xi=0$, $P_Z(p)=1+1/\{-\log(1-p)\}$.

(c) $Y=v+\sigma W$ is a location--scale transformation of $W$ with $\lambda=v/\sigma=\kappa^{-1}$, and~\eqref{eq:gpdPY} follows from~\eqref{eq:PlocScale} and~\eqref{eq:gpdMeanExcess}. Monotonicity in $\kappa$ follows from Proposition~\ref{prop:locScale}(ii), since $\lambda=\kappa^{-1}$ is decreasing in $\kappa$.

(d) If $0<\xi<1$, then $a_p\to\infty$ as $p\uparrow1$, so the second term in~\eqref{eq:gpdPZ} vanishes and $P_Z(p)\to1/(1-\xi)$; the same limit holds for $P_Y$ by Proposition~\ref{prop:locScale}(iv). If $\xi<0$, then $a_p\to0$, $a_p-1\to-1$ and $P_Z(p)\to1/(1-\xi)-\xi/(1-\xi)=1$; moreover $Q_W(p)\to-1/\xi<\infty$ and $e_W(p)\to0$, so $P_Y(p)\to1$. If $\xi=0$, $-\log(1-p)\to\infty$ gives $P_Z(p)\to1$ and $P_Y(p)=1+1/\{\kappa^{-1}-\log(1-p)\}\to1$.
\end{proof}

\subsection{Proof of Proposition~\ref{prop:gevPersistence}}
\label{app:proofGEV}

\begin{proof}
(a) For $\xi\ne0$, inverting~\eqref{eq:gevCDF} with $(\mu,\sigma)=(0,1)$ gives $Q_W(s)=(\ell_s^{-\xi}-1)/\xi$, $\ell_s=-\log s$. Substituting $w=-\log s$, so that $ds=-e^{-w}\,dw$, $s=p\mapsto w=\ell_p$ and $s=1\mapsto w=0$,
\[
\int_p^1(-\log s)^{-\xi}\,ds
=\int_0^{\ell_p} w^{(1-\xi)-1}e^{-w}\,dw
=\gamma(1-\xi,\ell_p),
\]
which is finite because $1-\xi>0$. Therefore
\[
m_W(p)
=\frac{1}{\xi}\left\{\frac{\gamma(1-\xi,\ell_p)}{1-p}-1\right\},
\qquad
e_W(p)=m_W(p)-z_p
=\frac{1}{\xi}\left\{\frac{\gamma(1-\xi,\ell_p)}{1-p}-\ell_p^{-\xi}\right\},
\]
and factoring out $\ell_p^{-\xi}$ yields the second form in~\eqref{eq:gevMeanExcess}. For $\xi=0$, $Q_W(s)=-\log(-\log s)$ and the stated expression is the definition~\eqref{eq:quantileMeanExcess}. Non-negativity of $e_W$ follows from Lemma~\ref{lem:meanExcess}.

(b) This is Proposition~\ref{prop:locScale} with $Q_W(p)=z_p$; strict monotonicity holds because $e_W(p)>0$ for the non-degenerate $W$.

(c) For $\xi\ne0$, $z_p=\omega_p-1/\xi$ and $e_W(p)=\omega_p\{G_\xi(p)-1\}$, so that $\lambda+z_p=c+\omega_p$ and
\[
P_Y(p)=1+\frac{\omega_p\{G_\xi(p)-1\}}{c+\omega_p}=\frac{c+\omega_p G_\xi(p)}{c+\omega_p}.
\]
Setting $c=0$ gives $P_Y(p)=G_\xi(p)$.

(d) As $p\uparrow1$, $\ell_p=-\log p\sim1-p\downarrow0$, and $\gamma(a,x)\sim x^a/a$ as $x\downarrow0$. Hence
\[
G_\xi(p)=\frac{\ell_p^{\,\xi}\,\gamma(1-\xi,\ell_p)}{1-p}
\sim\frac{\ell_p}{(1-\xi)(1-p)}
\longrightarrow\frac{1}{1-\xi}.
\]
If $0<\xi<1$, then $\omega_p\to\infty$, so by~\eqref{eq:gevWeighted} $P_Y(p)\to1/(1-\xi)$ for any fixed $c$, in agreement with Theorem~\ref{thm:persistence}. If $\xi<0$, then $W$ has finite upper endpoint $-1/\xi$, so $z_p\to-1/\xi$ and $e_W(p)\to0$, whence $P_Y(p)\to1$. If $\xi=0$, $z_p\to\infty$ while $e_W(p)\to1$, so $P_Y(p)\to1$.
\end{proof}

\subsection{Ordering of persistence across observations}
\label{app:proofContrast}

The following result justifies identity~\eqref{eq:persistenceContrast}.

\begin{proposition}
\label{prop:contrast}
Let $Y_i=\mu_i+\sigma_iW$ and $Y_{i'}=\mu_{i'}+\sigma_{i'}W$, where $W$ has a continuous, strictly increasing distribution depending only on $\xi$, and let $\lambda_i=\mu_i/\sigma_i$, $\lambda_{i'}=\mu_{i'}/\sigma_{i'}$. Fix $p\in(0,1)$ with $e_W(p)>0$, $Q_{Y_i}(p)>0$ and $Q_{Y_{i'}}(p)>0$. Then
\[
\{P_i(p)>P_{i'}(p)\}=\{\lambda_i<\lambda_{i'}\}
\]
as subsets of the parameter space. Consequently, for any posterior distribution of $(\mu_i,\sigma_i,\mu_{i'},\sigma_{i'},\xi)$ that gives probability one to these positivity conditions, $\Pr\{P_i(p)>P_{i'}(p)\mid\mathbf y\}=\Pr(\lambda_i<\lambda_{i'}\mid\mathbf y)$, and the common value does not depend on $p$.
\end{proposition}

\begin{proof}
By Proposition~\ref{prop:locScale}, $P_i(p)=g_p(\lambda_i)$ and $P_{i'}(p)=g_p(\lambda_{i'})$, with the \emph{same} function $g_p(\lambda)=1+e_W(p)/\{\lambda+Q_W(p)\}$, because $\xi$, and hence $W$, is common to $i$ and $i'$. The conditions $Q_{Y_i}(p)>0$ and $Q_{Y_{i'}}(p)>0$ are equivalent to $\lambda_i,\lambda_{i'}>-Q_W(p)$, and on this half-line $g_p$ is strictly decreasing by~\eqref{eq:dPdlambda}. Hence $g_p(\lambda_i)>g_p(\lambda_{i'})$ if and only if $\lambda_i<\lambda_{i'}$. The event on the right does not involve $p$, which proves the final statement. For the GPD level persistence the same argument applies with $\lambda_i=v_i/\sigma_i$ (Proposition~\ref{prop:gpdPersistence}(c)).
\end{proof}

\subsection{Persistence at unconditional levels of the parent distribution}
\label{app:uncond}

The following result justifies the remark following~\eqref{eq:gpdPYdraw}.

\begin{proposition}
\label{prop:uncond}
Let $Y$ have distribution $F$ with $\zeta=\Pr(Y>v)\in(0,1)$ for a threshold $v\ge0$, and suppose that $\Pr(Y-v>z\mid Y>v)=1-G_{\sigma,\xi}(z)$, $z\ge0$, with $\xi<1$. Then, for every $q\in[1-\zeta,1)$,
\[
P_F(q)=P_{Y}\!\left(1-\frac{1-q}{\zeta}\right),
\]
where $P_F$ is the persistence~\eqref{eq:persistence} of $F$ and $P_Y$ is the level persistence~\eqref{eq:gpdPY} of $v+Z$, $Z\sim GPD(\sigma,\xi)$.
\end{proposition}

\begin{proof}
For $z\ge0$, $\bar F(v+z)=\zeta\{1-G_{\sigma,\xi}(z)\}$. Let $p=1-(1-q)/\zeta\in[0,1)$. Then $\bar F\{v+Q_Z(p)\}=\zeta(1-p)=1-q$, so $Q_F(q)=v+Q_Z(p)$. Moreover, conditionally on $Y>Q_F(q)$, which implies $Y>v$, the distribution of $Y$ is that of $v+Z$ given $Z>Q_Z(p)$. Hence
\[
P_F(q)
=\frac{E\{Y\mid Y>Q_F(q)\}}{Q_F(q)}
=\frac{E\{v+Z\mid Z>Q_Z(p)\}}{v+Q_Z(p)}
=P_Y(p).
\qedhere
\]
\end{proof}

\end{document}